\documentclass[pra,amsmath, amssymb, superscriptaddress, showpacs, showkeys, notitlepage, longbibliography,twocolumn,nofootinbib]{revtex4-1}
\usepackage{graphicx}
\usepackage[usenames, dvipsnames]{color}
\usepackage{dsfont}
\usepackage[utf8]{inputenc}
\usepackage[english]{babel}
\usepackage{float}

\usepackage[breaklinks=true]{hyperref}
\hypersetup{
 colorlinks   = true, %Colours links instead of ugly boxes
 urlcolor     = blue, %Colour for external hyperlinks
 linkcolor    = blue, %Colour of internal links
 citecolor   =  red %Colour of citations
}
\usepackage{bbold}
\newcommand{\ket}[1]{\ensuremath{\left|#1\r\rangle}}

\renewcommand{\r}[0]{\right}

\newcommand{\Tr}{\text{Tr}}

\begin{document}
%\raggedbottom
\title{ Autonomous quantum heat engine using an electron shuttle }
\author{Behnam Tonekaboni} 
\email{uqbtonek@uq.edu.au}
\affiliation{School of Mathematics and Physics,  University of Queensland, Brisbane, QLD, 4072, Australia}
\affiliation{ARC Centre for Engineered Quantum Systems, The University of Queensland, Brisbane, QLD 4072, Australia}

\author{Brendon W. Lovett}
\affiliation{SUPA, School of Physics and Astronomy, University of St Andrews, St. Andrews KY16 9SS, United Kingdom}

\author{Thomas M. Stace}
\affiliation{School of Mathematics and Physics,  University of Queensland, Brisbane, QLD, 4072, Australia}
\affiliation{ARC Centre for Engineered Quantum Systems, The University of Queensland, Brisbane, QLD 4072, Australia}

\begin{abstract}
{We propose an autonomous quantum heat engine based on the thermally driven oscillation of a single electron shuttle. The electronic degree of freedom of this device acts as an internal dynamical controller which switches the interaction of the engine with the thermal baths. We show that in addition to energy flux through the thermal baths, a flux of energy is attributed to the controller and it affects the engine power.\\~\\~\\~\\~\\~}
\end{abstract}
\maketitle

%============================================================================================================
%                                                                                                     Introduction
%============================================================================================================
%\textit{Introduction}--- 
\section{Introduction}
With the advancement in experimental techniques, we are able to build and control devices with truly quantum degrees of freedom. In these quantum devices, the energetics and thermodynamics of the control field become important. Thermodynamics of such quantum devices have been studied in the context of quantum heat engines and fridges \cite{Kosloff-open, alicki1979quantum, Kosloff-Continuous}, driven nano-systems  \cite{Brandne-PeriodicTemp}, effect of energy transfer on quantum coherence \cite{scully2003extracting, scully2011quantum} and many other topics. Also a quantum heat engine has been built using a single ion \cite{Rossnagel-SingleAtomEngine}. The majority of these studies use an external time-dependent classical control field to switch the interaction between the working system and the heat baths. The energetic cost of these control fields is often ignored.
%Studies in quantum thermodynamics include different thermodynamics laws in systems with few degrees of freedom [??], the role of quantum information and coherence in thermodynamics [??] and quantum limit to the efficiency of thermal machines where achieving efficiency bigger than Carnot efficiency is open to debate.[??]\cite{Goold-quantinfothermo, Kosloff_dynamical}. 
%To analyse the efficiency in quantum regime, we need to design and construct thermal machines, specifically quantum heat engines \cite{lostaglio2015description}.  

Formally, the external classical control field is described by a time-dependent control parameter $ n_i \in [0,1] $ in the interaction Hamiltonian between the engine and baths,
%As in classical thermodynamics, quantum heat engines will be important for the potential of future quantum machines .
\begin{equation}
\label{eq:H_c_number}
H_{\mathrm{int}} = n_1(t) H_{\mathrm{int}}^{\mathrm{hot}} \otimes \mathbb{1}^{\mathrm{cold}} + n_2(t)  \mathbb{1}^{\mathrm{hot}} \otimes H_{\mathrm{int}}^{\mathrm{cold}},
\end{equation}
where $\mathbb{1}^{\mathrm{hot}}$ is the identity in Hilbert space of the hot bath and $H_{\mathrm{int}}^{\mathrm{hot}}$ is the interaction between working system and the hot bath, similar for $H_{\mathrm{int}}^{\mathrm{cold}}$. %A time-dependent Hamiltonian in the form of eq.\ref{eq:H_c_number} dismiss the effects of the controller on the engine. %In the macroscopic engines these effects are small but in a quantum heat engine care should be taken to account for this. 

%To be able to calculate the energy associated with the control field, we need to include the controller in to the total dynamics which leads to a time-independent Hamiltonian and therefore an autonomous heat engine. 

To account for the energetic cost of the control field we develop an autonomous heat engine with a time-independent Hamiltonian, in which the controller is included in system dynamics. Autonomous heat engines have received recent attention. \citet{Terry-QuasiAuto} studied a quasi-autonomous engine based on an internal quantum clock which is stabilised by external measurement. Although the interaction Hamiltonian used in \cite{Terry-QuasiAuto} is time independent, the external measurement result is used in a feedback loop to intermittently reset the clock. This induces a back action on the engine, so it is not a fully autonomous system. An autonomous rotor engine was proposed \cite{AutonomRotor} inspired by classical rotor engine where the state of the rotor determines the interactions and is considered as the internal clock. Neither of these papers address the energetic cost of the controller or the clock.

\begin{figure}[t]
\hspace*{-0.5cm}
\includegraphics[width= \columnwidth]{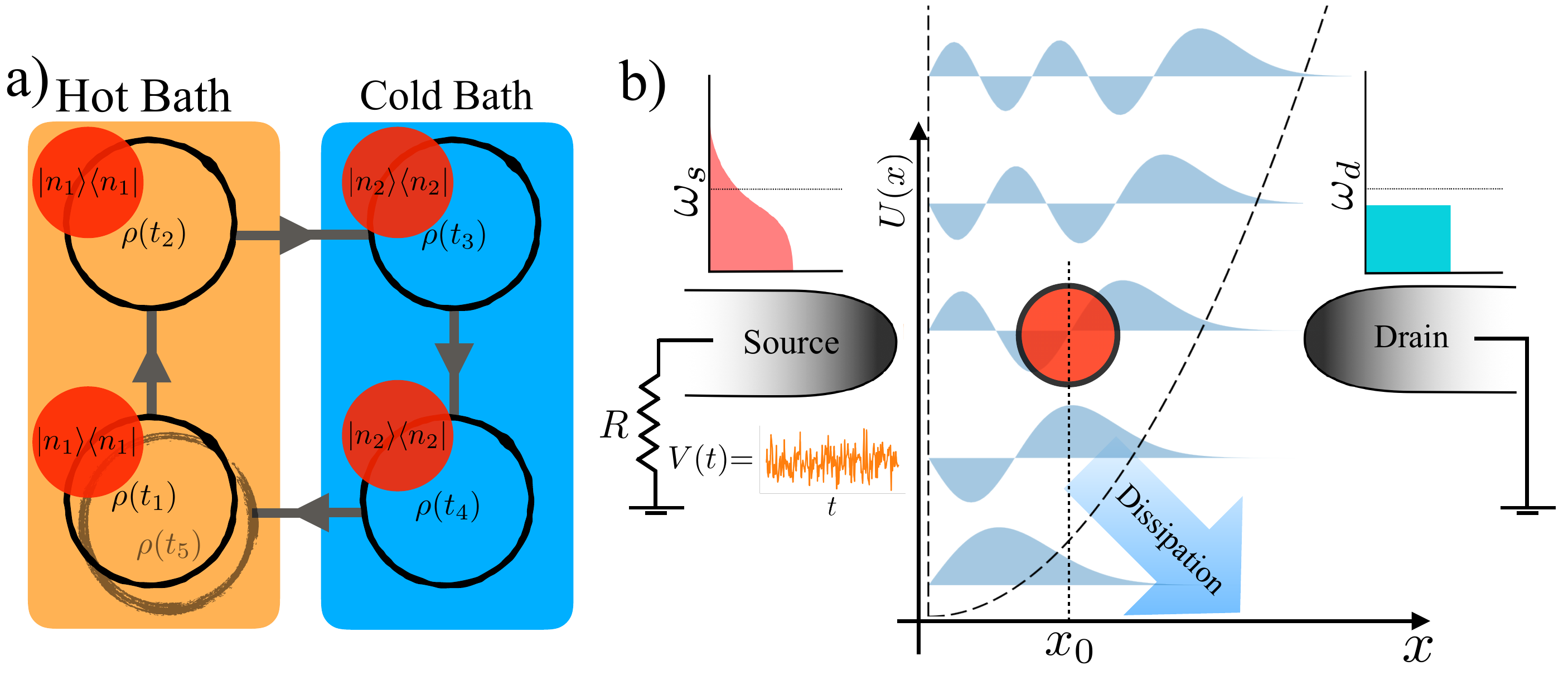}
\caption{a) shows thermodynamic cycle of the engine $\rho$ plus the controller $|n\rangle \langle n|$ between two thermal baths, hot and cold. The state of the engine changes due to thermal operations: $\rho(t_1) \rightarrow \rho(t_2)$ and $\rho(t_3) \rightarrow \rho(t_4)$ while it is in contact with hot and cold bath respectively. When the state of the controller changes, $|n_1\rangle \langle n_1| \rightarrow |n_2\rangle \langle n_2|$(or reverse), engine's interaction switches from hot to cold (or cold to hot). This switching changes the state of the engine $\rho(t_2) \rightarrow \rho(t_3)$(or $\rho(t_4) \rightarrow \rho(t_5)$). b) shows the diagram of the single electron shuttle heat engine. A single electron shuttle (red circle) is confined in a half-harmonic potential (dashed line), between two leads (source and drain). The leads have the same chemical potential but different temperatures, indicated by the Fermi distributions shown above them. Johnson noise from the thermal lead stochastically drives the charged island and the anharmonic potential rectifies it into a net force. The mechanical mode is coupled dissipatively to a cold bath.}
\label{fig:Diagrams}
\end{figure}

To fully account for the energetic cost of the controller, we develop a model of an autonomous engine, by replacing the external control field with an internal dynamical quantum controller that switches the interactions between the engine and the baths \cite{aaberg2018fully}. Formally we replace the control parameters in the interaction Hamiltonian with the orthonormal eigenstates of the internal dynamical controller $\ket{n_i}$ and write
\begin{equation}
\label{eq:H_control}
H_{\mathrm{int}} {=} |n_1\rangle \langle n_1 | \otimes  H_{\mathrm{int}}^{\mathrm{hot}} \otimes \mathbb{1}^{\mathrm{cold}} + |n_2\rangle \langle n_2 | \otimes \mathbb{1}^{\mathrm{hot}} \otimes H_{\mathrm{int}}^{\mathrm{cold}}.
\end{equation}
Figure \ref{fig:Diagrams}a shows a schematic of a thermodynamic cycle using a dynamical controller. When the controller is in the state $\ket{n_1}$ the engine interacts with the hot bath and when it is in the state $\ket{n_2}$ the engine interacts with the cold bath. 

A heat engine absorbs heat from a hot bath to increase the energy of the working system. In quantum thermodynamics, this process happens in a \emph{thermal operation} \cite{horodecki2013fundamental, sparaciari2017resource, Goold-quantinfothermo, aaberg2014catalytic}. One condition on the thermal operation is that the pure state of any ancilla (catalyst or in our case the controller) attached to the system must be the same at the beginning and the end of the thermal operation. On the other hand, in an autonomous heat engine, the controller has its own dynamics, and stochastic fluctuations during each cycle make it difficult to unambiguously define discrete phases of thermal operations. %and changes its state to be able to switch the interactions. Therefore it is not purely thermal operation and care should be taken in energy analysis.

In this paper, we introduce a fully autonomous quantum heat engine based on the oscillation of a single-electron shuttle  \cite{MechanicalShuttle,Gerard-shuttle}, in which the charge state of the shuttle acts as the control system coupling the engine to the hot bath. In section II we describe a model for this system and define the engine and the controller. In order to identify the heat flow from the hot and the cold reservoirs, in section III we briefly describe a simpler system, a charged thermally driven oscillator, where the state of the controller is held constant.  In section IV we include the dynamical controller and let it evolve during the thermal operation to see its effect on the internal energy of the engine. Also we analyse the energy flow in an unravelling of the system master equation, to derive a system of stochastic dynamic equations for the engine. In section V we numerically solve and analyse dynamics of the engine. At the end, in section VI, we calculate power of the engine in semiclassical and quantum approaches and show that quantum correlations has an important role in the engine power.
%============================================================================================================
%                                                                                                           Model
%============================================================================================================
%\textit{Model}--- 
\section{Model}
Our proposed engine is a modified version of the electromechanical shuttle studied in \cite{Gerard-shuttle}. Figure \ref{fig:Diagrams}b illustrates the engine, consisting of a single electron island that mechanically oscillates in a potential between two leads, shuttling an electron from the source to the drain over each cycle. In \cite{Gerard-shuttle} the voltage bias between the leads generates an electric field that drives the charged shuttle, so that it is a microscopic electric motor. In contrast, we drive the shuttle with a thermal bias (at zero voltage bias), corresponding to a microscopic thermal engine. We set the leads to the same chemical potential while allowing them to be at finite temperatures. Johnson noise from the finite temperature leads provides a stochastic driving force.  

Johnson noise is approximated as white noise with zero mean, so the time averaged force is zero. To rectify this noise into a net force, we impose a nonlinear potential. For this theoretical model we use a half-harmonic potential (shown with dashed line in Figure \ref{fig:Diagrams}b) which is easy to analyse and provides the required nonlinearity. The energy spectrum of the half-harmonic oscillator is equally spaced with an energy gap of $2\omega$, which is described by a simple Hamiltonian, $H_{\text{osc}}=2\omega a^\dagger a$ where $a$ is the bosonic annihilation operator of the oscillator and satisfies $[a,a^\dagger]=1$. Although the Hamiltonian of the half-harmonic oscillator is simillar to the full-harmonic, the position and momentum operators are not the same as full-harmonic due to the asymmetry in eigen-wavefunctions which are illustrated by blue shaded area in Figure \ref{fig:Diagrams}b (For details see Appendix B). This asymmetry provides larger $\langle \hat{\mathrm{x}} \rangle$ for higher energy. Therefore, when Johnson noise adds energy, the oscillator feels a rectified force to the right (For details see Appendix C).

Finally, the oscillator is coupled to a cold bosonic environment, which provides a cold bath and a sink for the work output of the engine.

In our model we assume the total Hamiltonian of the shuttle, the leads, Johnson noise and the bosonic environment is
\begin{subequations}
\label{eq:Hamiltonian}
\begin{align}
\hat{H} &= \hbar \omega_I \hat{c}^\dagger \hat{c} \label{subeq:electron} \\ %+ U_c \hat{c}^\dagger \hat{c}  \\
            &+ \sum_k \hbar \omega_{sk} \hat{b}_{sk}^\dagger \hat{b}_{sk} + \hbar \omega_{dk} \hat{b}_{dk}^\dagger \hat{b}_{dk}  \label{subeq:leads}\\
            &+ \sum_k \left[ \tau_{sk} F_s(\hat{\mathrm{x}}) \hat{b}_{sk} \hat{c}^\dagger + \text{H.c.} \right] \label{subeq:tunnel_s}\\
            &+ \sum_k \left[ \tau_{dk} F_d(\hat{\mathrm{x}}) \hat{b}_{dk} \hat{c}^\dagger + \text{H.c.} \right] \label{subeq:tunnel_d} \\
            &+ 2 \hbar \omega \hat{a}^\dagger \hat{a} \label{subeq:oscillator} \\ 
            &- e \xi(t) \hat{\mathrm{x}} \hat{c}^\dagger \hat{c} \label{subeq:Johnson}\\
            &+ \sum_p g(\hat{a}^\dagger \hat{d}_p +\hat{a} \hat{d}_p^\dagger) + \sum_p \hbar \omega_p \hat{d}_p^\dagger \hat{d}_p \label{subeq:bosonic}.
\end{align}
\end{subequations}
Terms  (\ref{subeq:electron}) to (\ref{subeq:tunnel_d}) describe dynamics of the controller. Terms (\ref{subeq:electron}) is the single electron energy where $\omega_I$ is the electronic energy level of the shuttle. We assume the island charging energy is large so that we retain only a single electronic mode with fermionic annihilation operator $\hat{c}$ which satisfies $\{\hat{c},\hat{c}^\dagger\} {=} 1$. The Hilbert space of the electrical degree of freedom is then $\{|0\rangle, |1 \rangle\}$, where $|n_e\rangle$ is the state of $n_e$ electron on the shuttle.   %The second term in (\ref{subeq:electron}) is Coulomb blockade [?], where we  assume the charging energy $U_c$ is large enough that no more than one electron occupies the shuttle. 

The Hamiltonian of the leads is given in  (\ref{subeq:leads}), where $\hat{b}_{s(d),k}$ is fermionic annihilation operator for the source(drain) at mode $k$ with energy $\hbar \omega_{s(d)k}$. Terms (\ref{subeq:tunnel_s}) and (\ref{subeq:tunnel_d}) are electron tunneling terms between the leads and the shuttle.  The tunneling functions $F_{s(d)}$ indicate the position dependency of the tunneling terms, and $\tau_{s(d),k}$ is the tunnelling constant between the shuttle and mode $k$ of the source(drain). 

Expressions (\ref{subeq:oscillator}) to (\ref{subeq:bosonic}) describe the mechanical degree of freedom. Expression (\ref{subeq:oscillator}) is the Hamiltonian of the oscillator in the half-harmonic potential while (\ref{subeq:Johnson}) is the coupling between thermal (Johnson) noise $\xi$  and the shuttle position $\hat{\mathrm{x}}$. We assume $\langle \xi(t) \rangle_\text{noise} = 0$ and $\langle \xi(t) \xi(t') \rangle_\text{noise}=E_{\mathrm{rms}}^2 \delta(t-t')$, where $E_{\mathrm{rms}}=4k_BTR$ is the noise amplitude with temperature $T$ of the source lead and $R$ is the lead resistance. This term can be written as $-e\xi \hat{\mathrm{x}} |1\rangle \langle 1|$, which is in the form of equation (\ref{eq:H_control}). The shuttle  couples to the thermal noise (hot bath) when an electron is on board (i.e. the electronic state is $|1\rangle$), and is uncoupled when it is unoccupied (i.e. the electronic state is $|0\rangle$).

The last line in the Hamiltonian, (\ref{subeq:bosonic}), describes the interaction of the oscillator with the bosonic environment where $\hat{d}_p$ is the bosonic annihilation operator of mode $p$ with energy $\hbar \omega_p$, while $g$ is a coupling constant. % We would like to clarify that in this system, the mechanical degree of freedom (i.e. oscillator) is the engine and the electrical degree of freedom (i.e. presence of an electron on the shuttle) is the controller.

We obtain a master equation for the shuttle by tracing out the leads and bosonic reservoir and averaging over the noise. We define the reduceed density operator for the shuttle, following averaging over Johnson noise as $\rho \equiv \langle \Tr_{L,B}\{\rho_T\} \rangle_{\text{noise}}$, where $\rho_T$ is the density operator of the total system. In the Born-Markov approximation \cite{gardiner2004quantum} the master equation is (See appendix D for derivation)
\begin{align}
\label{eq:master_full}
\dot{\rho} =  &-i \left[ 2 \omega \hat{a}^\dagger \hat{a} , \rho \right] + \gamma \mathcal{L}[\hat{c}^\dagger \hat{c} \hat{\mathrm{x}}] \rho \\
                    &+ \kappa \left( \overline{n}_p + 1\right) \mathcal{L}[\hat{a}]\rho + \kappa \overline{n}_p \mathcal{L}[\hat{a}^\dagger]\rho \nonumber \\
                    &+ \Gamma_s \left\{ f_s(\omega_I) \mathcal{L}[\hat{c}^\dagger F_s(\hat{\mathrm{x}}) ]\rho + \left( 1 {-}  f_s(\omega_I) \right) \mathcal{L}[\hat{c} F_s(\hat{\mathrm{x}}) ]\rho  \right\} \nonumber\\
                    &+ \Gamma_d \left\{ f_d(\omega_I) \mathcal{L}[\hat{c}^\dagger F_d(\hat{\mathrm{x}}) ]\rho + \left( 1 {-}  f_d(\omega_I) \right) \mathcal{L}[\hat{c} F_d(\hat{\mathrm{x}}) ]\rho  \right\}, \nonumber
\end{align}
where $\mathcal{L}[O]\rho = O\rho O^\dagger -\frac{1}{2} [O^\dagger O \rho + \rho O^\dagger O]$ is the Lindblad super-operator \cite{lindblad1976generators}. Also we use dimensionless units where $\hbar=1$. 

The first two lines of equation (\ref{eq:master_full}) describe the dynamics of the engine. The first term gives the evolution in the half-harmonic potential while the second term is an incoherent driving term due to Johnson noise with rate $\gamma = e^2 E^2_{rms} \propto T^2$. Here we have assumed that Johnson noise is fast compared to the oscillator dynamics. As a result, the stochastic driving field appears as a dissipative term (See Appendix A and \cite{Stace-StochNoise})\footnote{In the voltage driven shuttle in \cite{Gerard-shuttle}, a time-independent electric field acts on the shuttle, thus the driving term is coherent and appears as  $i[\hat{c}^\dagger \hat{c} \hat{\mathrm{x}},\rho]$. This is in contrast to the dissipative term in equation (\ref{eq:master_full}).}. The second line of the master equation is the damping due to the cold bath with the mean phonon number $\overline{n}_p = 1/(e^{\omega/T} - 1)$ and damping rate $\kappa$. The last two lines of equation (\ref{eq:master_full}) describe the electron tunnelling between the leads and the shuttle which controls the engine--bath coupling. The rate of tunnelling depends on the Fermi distribution, $f_{s(d)}(\omega_I) = 1/(e^{(\omega_I - \mu)/T_{s(d)}}+1)$ where $T_{s(d)}$ is temperature of the source(drain) and $\mu$ is the common chemical potential.

We choose $\omega_I$ slightly greater than the chemical potential of the leads. Thus at zero temperature electrons are energetically excluded from the shuttle if we neglect cotunnelling processes. At finite temperature, electrons can jump on and off the shuttle. We set the drain temperature to be small so that we can approximate $f_d (\omega_I)\approx 0$. This ensures electrons do not tunnel from the drain to the shuttle and therefore the shuttle carries an electron in each cycle from the source to the drain. 

The leads are positioned so that the separation is comparable to the oscillator amplitude, therefore the tunnelling amplitude depends on the shuttle position. We model this by defining $F_s(\hat{\mathrm{x}}) =  \alpha_s e^{-\eta \hat{\mathrm{x}}}$ and $F_d(\hat{\mathrm{x}}) =  \alpha_d e^{\eta \hat{\mathrm{x}}}$, where $\eta$ is the inverse of the tunnelling length and $\alpha_s {=} A e^{\eta x_0}$ and  $\alpha_d {=} A e^{-\eta x_0}$, where $x_0$ is the midpoint between the source and the drain, as shown in figure \ref{fig:Diagrams}b.

For the purpose of this paper we study the dynamics of the shuttle assuming the temperature of the cold bosonic bath is low and $\overline{n}_p$ is near zero (For more general thermodynamic analysis this assumption can be lifted, this will be the subject of the future work). The master equation then becomes
\begin{align}
\label{eq:master}
\dot{\rho} =  &-i \left[ 2 \omega \hat{a}^\dagger \hat{a} , \rho \right] + \gamma \mathcal{L}[\hat{c}^\dagger \hat{c} \hat{\mathrm{x}}] \rho + \kappa \mathcal{L}[\hat{a}]\rho \nonumber \\
                    &+ \Gamma_s \left\{ f_s(\omega_I) \alpha_s^2 \mathcal{L}[\hat{c}^\dagger e^{-\eta \hat{\mathrm{x}}} ]\rho + \left( 1 -  f_s(\omega_I) \right) \alpha_s^2 \mathcal{L}[\hat{c} e^{-\eta \hat{\textrm{x}}} ]\rho  \right\} \nonumber \\
                    &+ \Gamma_d \alpha_d^2 \mathcal{L}[\hat{c} e^{\eta \hat{\textrm{x}}} ]\rho.
\end{align}

Since the engine is the mechanical degree of freedom of the shuttle, the energy of the working system is the energy of the oscillator $E{=}\langle H_\textrm{osc} \rangle {=} 2 \omega \langle \hat{a}^\dagger \hat{a} \rangle$. The rate of the change of the mechanical energy of the oscillator is 
\begin{align}
\frac{dE}{dt} = \Tr[H_\mathrm{osc} \dot{\rho}].
\label{eq:energyrate}
\end{align}

In the following we analyse this energy rate using a fixed charge (in section III) and full dynamics with electron jumps (in section IV).
%============================================================================================================
%                                                                                                     Charged Oscillator 
%============================================================================================================
\medbreak
%\textit{Charged Oscillator}--- 
\section{Charged Oscillator}
The charge state of the shuttle determines the Johnson noise coupling. As a prelude to studying the full dynamics, we analyze the evolution for a fixed charge state  of the shuttle. By using a fixed controller state (fixed charge) we focus to the thermal operation and identify heat flow to and from the thermal baths. To model this, we set a fixed charge $\hat{c}^\dagger \hat{c} = n_e$ on the shuttle, and set $\alpha_{s,d}=0$. The master equation for the fixed charge shuttle is then
\begin{equation}
\label{eq:master_charged}
\dot{\rho} = -i \left[2\omega \hat{a}^\dagger \hat{a}, \rho \right]  + \gamma n_e^2 \mathcal{L}[\hat{\mathrm{x}}]\rho + \kappa \mathcal{L}[\hat{a}]\rho.
\end{equation}

We calculate the rate of change of energy in the oscillator, i.e. equation (\ref{eq:energyrate}), during the thermal operation, and we find
\begin{align}
\frac{dE}{dt} %&= \Tr[H_\mathrm{osc} \dot{\rho}] \nonumber \\
                    %&= 2 \omega \Tr [a^\dagger a \dot{\rho}] \nonumber \\
                    &=\frac{\omega n_e^2 \gamma }{2}  - 2 \omega \kappa N  \label{eq:energy_charged} \\
                    &\equiv \dot{\mathcal{Q}}_\text{H} - \dot{\mathcal{Q}}_\text{C} \label{eq:thermo},
\end{align}
where $N=\langle a^\dagger a \rangle$ is the expectation value of phonon number. In equation (\ref{eq:thermo}) we have defined the heat flux from the hot bath (i.e. Johnson noise) $\dot{\mathcal{Q}}_\text{H} = \frac {\omega n_e^2 \gamma }{2}$, and the energy flux to the cold bath $\dot{\mathcal{Q}}_\text{C} = 2 \omega \kappa N$. When $n_e=1$, the oscillator couples to Johnson noise which causes a steady increase in the mechanical energy of the system and when $n_e=0$ the oscillator is only in contact with the bosonic cold bath. We interpret equation (\ref{eq:thermo}) as a thermodynamic identity of the thermal operation with a static control parameter. 
%============================================================================================================
%                                                                                              Autonomous Engine
%============================================================================================================
\section{Autonomous Engine}
Equation (\ref{eq:energy_charged}) and its thermodynamic interpretation equation (\ref{eq:thermo}) consider only the energy flow from and to the thermal baths. They do not include power attributed to the dynamical controller (i.e. tunnelling). To account for this, we calculate energy rate, equation (\ref{eq:energyrate}), from the full dynamical system, including the controller i.e. equation (\ref{eq:master}). We find
%To include this we calculate $\dot{E} = 2 \omega \Tr [a^\dagger a \dot{\rho} ]$ where we substitute $\dot{\rho}$ with equation (\ref{eq:master}) and 

\begin{align}
\frac{dE}{dt} &=  \frac{\omega \gamma  \langle c^\dagger c \rangle^2 }{2} - 2 \omega \kappa N \nonumber \\
                    &+ \frac{1}{2} \omega \Gamma_s \alpha_s^2 \eta^2 \left[ f_s(1- \langle c^\dagger c \rangle) + (1- f_s)  \langle c^\dagger c \rangle \right] \langle e^{-2\eta \hat{\mathrm{x}}} \rangle \nonumber \\
                    &+ \frac{1}{2} \omega \Gamma_d \alpha_d^2 \eta^2  \langle c^\dagger c \rangle \langle e^{2\eta \hat{\mathrm{x}}} \rangle \label{eq:energy}\\
                    &\equiv \dot{\mathcal{Q}}_\text{H} - \dot{\mathcal{Q}}_\text{C} + \dot{E}_{\mathrm{control}}. \label{eq:thermo_identity}
\end{align}
where $\dot{E}_\mathrm{control}$ 
%which is given by the last terms of equation (\ref{eq:energy}),
is the contribution of the controller to the energy of the engine which can be positive or negative, meaning that the controller can add or take energy from the oscillator. Thus the thermodynamic identity for the whole system must include the control power as well as the thermal operation.

To describe the stochastic evolution of the system, we unravel the master equation into a stochastic master equation (SME), using a jump process for the charge state of the shuttle. In this unravelling the controller charge state fluctuates stochastically between $|0\rangle$ and $|1\rangle$.
In an experimental setup, the SME has an interpretation as a model of continuous measurement, in which one monitors the electron jumps and thus the evolution of the controller \cite{goan2001continuous}. The SME for the conditional state of the engine is
\begin{widetext}
\begin{subequations}\label{eq:conditional_master}
\begin{align} 
d \rho_c =  &-i \left[ 2 \omega \hat{a}^\dagger \hat{a} , \rho_c \right]dt + \gamma \mathcal{L}[\hat{c}^\dagger \hat{c} \hat{\textrm{x}}] \rho_c dt + \kappa \mathcal{L}[\hat{a}]\rho_c \label{subeq:conditional_unconditional} dt \\
                     &-\frac12 dt \left\{ \Gamma_s f_s(\omega_I) \alpha_s^2  \mathcal{H}[\hat{c} \hat{c}^\dagger e^{-2\eta \hat{\mathrm{x}}}]\rho_c +  \Gamma_s(1- f_s(\omega_I)) \alpha_s^2  \mathcal{H}[\hat{c}^\dagger \hat{c} e^{-2\eta \hat{\mathrm{x}}}]\rho_c \right\} 
                    -\frac12 dt~  \Gamma_d \alpha_d^2  \mathcal{H}[\hat{c}^\dagger \hat{c}  e^{2\eta \hat{\mathrm{x}}}]\rho_c \label{subeq:conditional_measurement}\\
                    &+dN_{+}^s \mathcal{G}[\hat{c}^\dagger e^{-\eta \hat{\mathrm{x}}}]\rho_c +dN_{-}^s \mathcal{G}[\hat{c} e^{-\eta \hat{\mathrm{x}}}]\rho_c + dN_{-}^d \mathcal{G}[\hat{c} e^{\eta \hat{\mathrm{x}}}]\rho_c \label{subeq:conditional_jump},
\end{align}
\end{subequations}
\end{widetext}
where $\mathcal{G}$ and $\mathcal{H}$ are nonlinear superoperators defined by
\begin{align}
\mathcal{G}[O]\rho &=\frac{O\rho O^\dagger}{\Tr[O\rho O^\dagger]} - \rho \\
\mathcal{H}[O]\rho &=O \rho + \rho O^\dagger - \Tr[O \rho + \rho O^\dagger]\rho.
\end{align}
Terms in (\ref{subeq:conditional_unconditional}) are the same as the first line of the equation (\ref{eq:master}) while (\ref{subeq:conditional_jump}) describes electron jumps on and off the shuttle with stochastic increments $dN_\pm^s$ and $dN_-^d$ which are either $0$ or $1$. The probabilities of these stochastic terms depend on the state of the shuttle and have the following mean values:
\begin{subequations}
\begin{align}
\langle dN_{+}^s \rangle &= \Gamma_s f_s(\omega_I) \alpha_s^2 \Tr[\hat{c} \hat{c}^\dagger e^{-2\eta \hat{\textrm{x}}}\rho_c] dt\\
\langle dN_{-}^s \rangle &=  \Gamma_s(1- f_s(\omega_I)) \alpha_s^2  \Tr[\hat{c}^\dagger \hat{c} e^{-2\eta \hat{\mathrm{x}}}\rho_c] dt\\
\langle dN_{-}^d \rangle &= \Gamma_d \alpha_d^2  \Tr[\hat{c}^\dagger \hat{c} e^{2\eta \hat{\mathrm{x}}} \rho_c] dt .
\end{align}
\end{subequations}

In the SME (\ref{eq:conditional_master}), the controller has two energetic effects: 1) Back action of continuous current measurement, 2) Electron jumps which switch the bath interactions. To show this we write the change in the mechanical energy of the oscillator as
\begin{align}
dE_c &= \Tr[H_\mathrm{osc} d \rho_c] \nonumber \\
         &= (\dot{\mathcal{Q}}_\text{H} - \dot{\mathcal{Q}}_\text{C} + \dot{E}_M) dt + dE_J
\end{align}
where $\dot{\mathcal{Q}}_\text{H}$ and $\dot{\mathcal{Q}}_\text{C}$ are as before while $\dot{E}_M dt = \Tr[H_\mathrm{osc} \{ \text{expression } \ref{subeq:conditional_measurement} \}]$ describes the energy transfer due to continuous current measurement; and $dE_J = \Tr[H_\mathrm{osc} \{ \text{expression } \ref{subeq:conditional_jump} \}]$ is a stochastic term due to the electron jumps on and off the shuttle.

\begin{figure}[t!]
\includegraphics[width=\columnwidth]{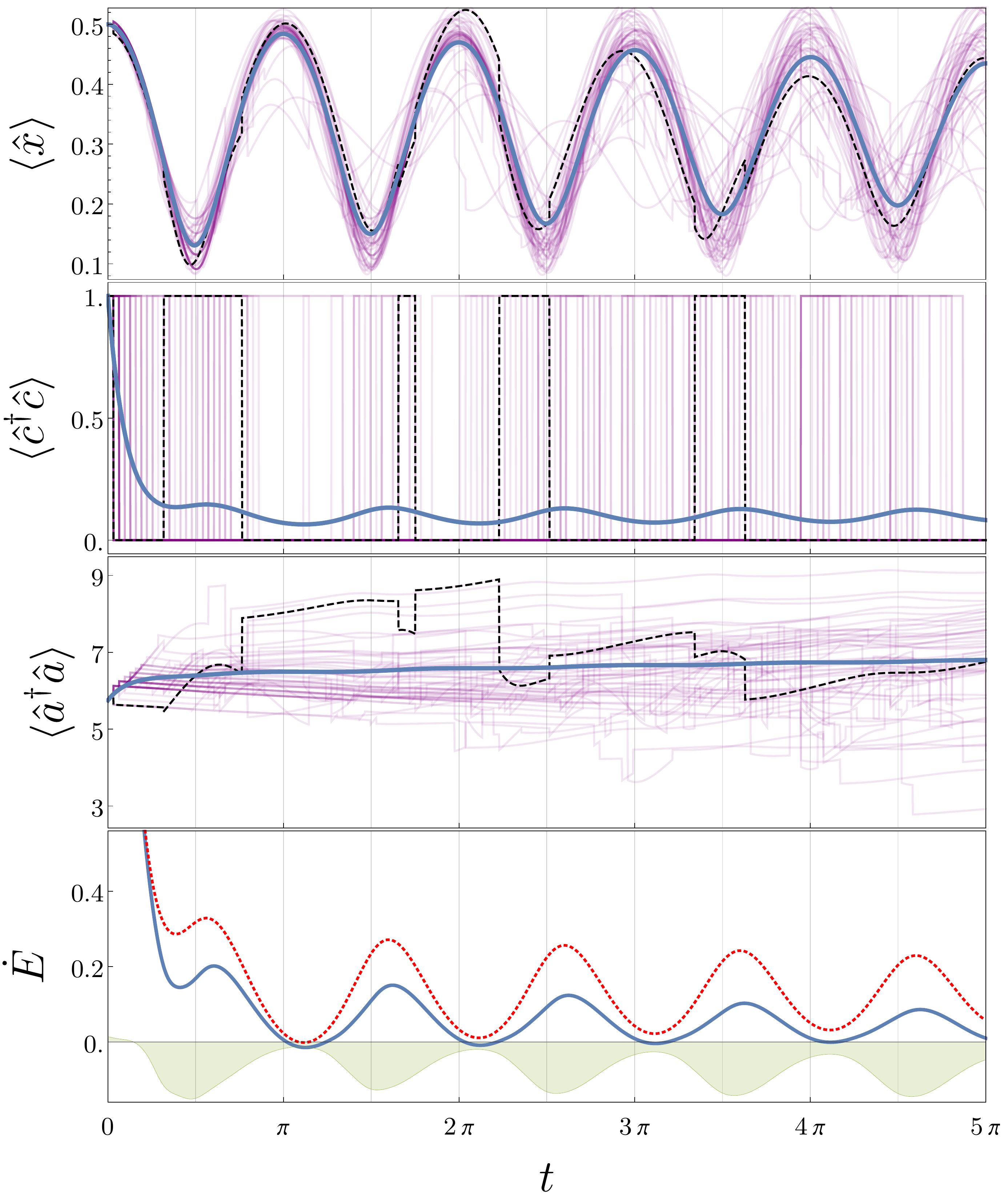}
\caption{Shows the time evolution of the expectation value of (a) position $\langle \hat{\mathrm{x}} \rangle$, (b) electron number $\langle c^\dagger c \rangle$ and, (c) phonon number $\langle a^\dagger a \rangle$. The solid blue lines show the unconditional solution to the master equation (\ref{eq:master}). Light purple lines are 500 trajectories of conditional dynamics, equation (\ref{eq:conditional_master}), indicating the typical spread, and the black dashed line highlights a single trajectory. (d) shows the energy flux through the oscillator (blue solid line), flux due to thermodynamic reservoirs (red dotted line), and flux due to the dynamical controller $\dot{E}_{\mathrm{control}}$ (shaded green)  }
\label{fig:time_evolution}
\end{figure}
%============================================================================================================
%                                                                                               Numerical Solution
%============================================================================================================
%\textit{Numerical Solution}--- 
\section{Dynamical simulations}

In this section we analyze the behaviour of the engine and the effect of the controller by solving the unconditional master equation (\ref{eq:master}) and SME (\ref{eq:conditional_master}). 

We use a truncated oscillator basis and find the time evolution of the matrix elements of $\rho$ in the eigenbasis of the oscillator and the electrons. Then we calculate the expectation values of the position $\langle \hat{\mathrm{x}} \rangle$, the phonon number $\langle \hat{a}^\dagger \hat{a} \rangle$ and the electron number $\langle \hat{c}^\dagger \hat{c} \rangle$.  The solid blue curves in figures \ref{fig:time_evolution}a--c show these quantities as functions of time. Figures \ref{fig:time_evolution}a and  \ref{fig:time_evolution}b show the shuttling behaviour of the engine: when the shuttle is near to the source lead ($x_\text{source} = 0$), average number of electrons, $\langle \hat{c}^\dagger \hat{c} \rangle$, increases and when $\langle \hat{\mathrm{x}} \rangle$ is near to the drain  ($x_\text{drain} = 0.6$) the electron number decreases. Figure \ref{fig:time_evolution}c shows $\langle \hat{a}^\dagger \hat{a} \rangle$ and consequently mechanical energy of the engine $E=2\omega \langle \hat{a}^\dagger \hat{a} \rangle$ increases. 

Figure \ref{fig:time_evolution}d shows the rate of change in mechanical energy of the engine. Peaks in $\dot{E}$ match with the peaks in $\langle \hat{c}^\dagger \hat{c} \rangle$ and it confirms that the presence of an electron acts as the controller: when the shuttle is charged the heat flux to the engine is the largest. The net heat flux $\dot{\mathcal{Q}}_H - \dot{\mathcal{Q}}_C$ to the engine is plotted as a dotted red line while the green shaded area shows the energetic cost of the control power. The magnitude of the controller effect increases while electron number increases (and the shuttle is near to the source). This is consistent with equation (\ref{eq:energy}) where electron number, $ \langle c^\dagger c \rangle$, has an important role in $\dot{E}_\text{control}$.

A solution to the SME (\ref{eq:conditional_master}) is a quantum trajectory. The light purple lines in figures \ref{fig:time_evolution}a-c show 500 trajectories; one is highlighted in black dashed line. The spread of trajectories indicates the stochastic variation of $\hat{\mathrm{x}}$, $\hat{c}^\dagger \hat{c}$ and $\hat{a}^\dagger \hat{a}$ in figure \ref{fig:time_evolution}a-c. 

\begin{figure}[t]
\centering
\includegraphics[width=\columnwidth]{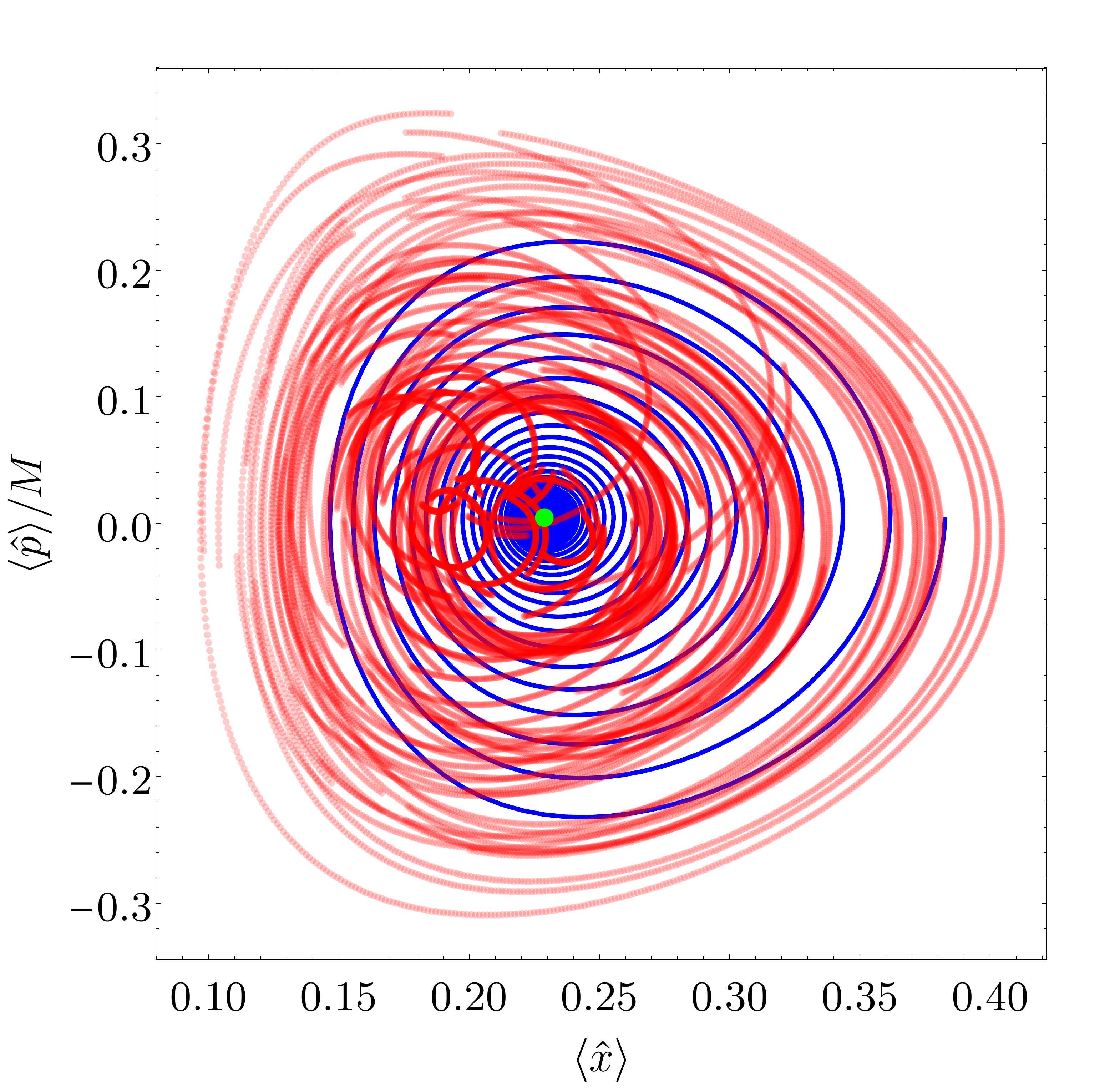}
\caption{Shows the phase space diagram of the oscillator coordinates $\text{v}=\langle \hat{\mathrm{p}} \rangle/m$ and $\langle \hat{\mathrm{x}} \rangle$. Solid blue line shows the unconditional coordinates up to $t= 30 \pi$ (i.e. 30 cycles), which is substantially longer than $1/\kappa=20$ representing the damping time scale for the unconditional dynamics.  Red dots show 50 cycles of a typical stochastic trajectory from $t=450\pi$ to $t=500\pi$, i.e.\ in the long time limit. This illustrates the fact that typical trajectories do not dwell near the unconditional steady state (green point).}
\label{fig:phasediagram}
\end{figure}

We plot the oscillator phase space coordinates $v=\langle \hat{\mathrm{p}} \rangle/M$ and $\langle \hat{\mathrm{x}} \rangle$, where $M$ is the mass of the oscillator. The solid blue line in figure 3 shows the unconditional coordinates over 50 cycles. Phase fluctuations in the cycles cause the oscillator to damp to an unconditional steady state.  

In contrast, a typical trajectory in the stochastic solution to the SME does not converge to a well defined fixed point. Instead it continues to fluctuate around the unconditional fixed point, rarely dwelling near it. A representative trajectory is shown as red points in Figure \ref{fig:phasediagram}, where we have plotted the last 50 cycles to show that the unconditional steady state is not representative of  the long time stochastic dynamics; that is, stochastic fluctuations remain large in the long time dynamics.

% By looking at a single trajectory in the phase diagram, we see the shuttle oscillates stochastically without general damping. The red dots in figure 3 shows a single trajectory for 500 cycles. The initial state of this trajectory is chosen the same as the unconditional solution (Blue line). We observe that the amplitude of the stochastic trajectory sometimes goes beyond the the initial amplitude.

\begin{figure}[t!]
\centering
\includegraphics[width=\columnwidth]{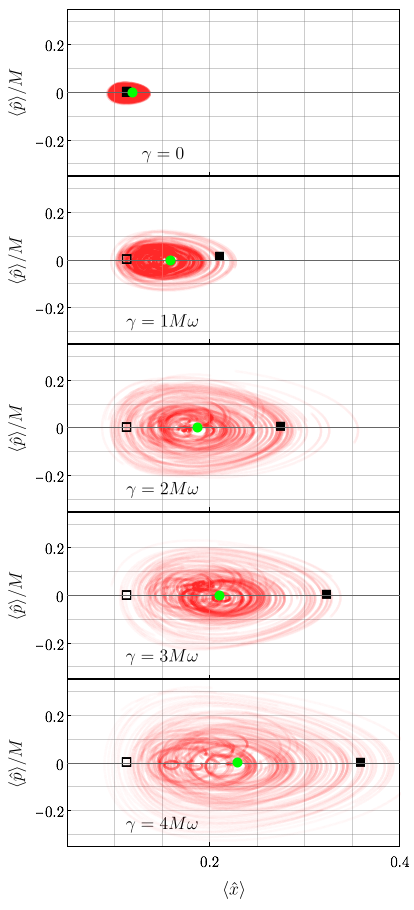}
\caption{shows phase diagrams of typical trajectories of the oscillator for different values of $\gamma$ and fixed damping rate $\kappa=0.05$. The unconditional steady states are shown by green points.
The hollow and the filled squares are fixed points for uncharged, $n_e=0$, and charged,  $n_e=1$, oscillator respectively. For $\gamma=0$ a typical trajectory fluctuates randomly around the steady state within the size of the zero point motion $1/\sqrt{M\omega}$. As $\gamma$ increases the the stochastic cycles sweep a larger area.}
\label{fig:limitcycle}
\end{figure}

We expect that the swept phase space area grows as the driving strength increases. This is demonstrated in Figure \ref{fig:limitcycle}, which shows a series of stochastic solutions to the SME (\ref{eq:conditional_master}) as the strength of the coupling to the Johnson noise increases. Clearly, the phase space region explored by the shuttle grows as the coupling (and heat flux) increases.

\begin{figure}[t!]
\centering
\includegraphics[width=\columnwidth]{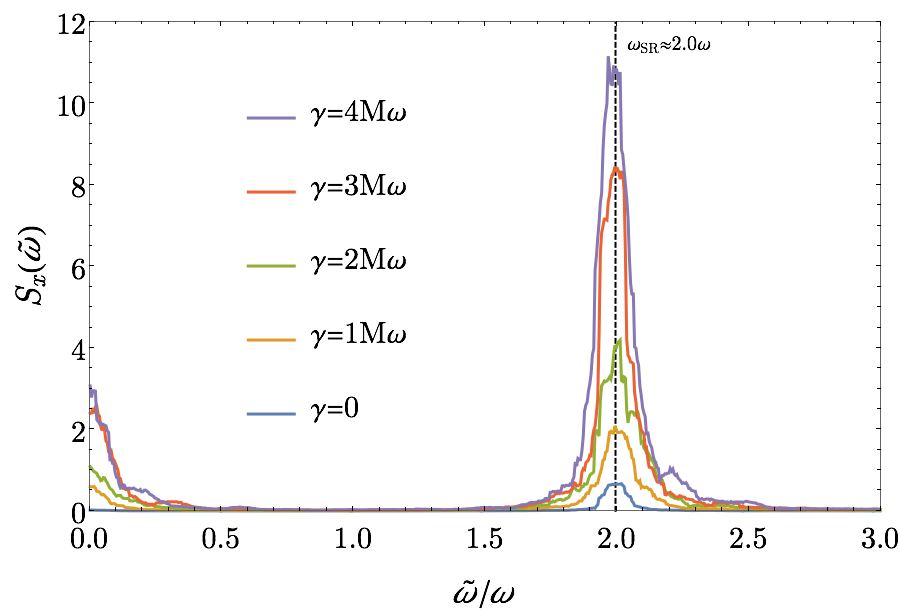}
\caption{shows power spectrum of oscillator position for different values of $\gamma$. The peak at $\omega_{SR}\approx2.0\omega$ indicates the existence of stochastic resonance.}
\label{fig:powerspectrum}
\end{figure}

\emph{Discussion--} Nonlinear systems subject to noise exhibit various phenomenon including stochastic limit cycle \cite{itoh1994stochastic} and stochastic resonance \cite{StochasticRes}. Stochastic resonance (SR) in a nonlinear system occurs when a time-dependent, periodic force is amplified by  noise, and arises from a the emergence of stochastic limit cycle, in which two stable attractors become coupled by the noise.  %behaviour in a non-linear system subject to periodic and noisy forces. 
Since  the  discovery of SR, \citet{gang1993stochastic} and others \cite{rappel1994stochastic} showed that nonlinear systems with fixed points subject to noise can exhibit \emph{autonomous} stochastic resonance (ASR) i.e.\ without an externally applied periodic force. 

The system we consider in this paper falls into this category: the quantum shuttle has two fixed points associated to each of the charge states of the shuttle, i.e.\ for $n_e=\langle \hat{c}^\dagger \hat{c} \rangle=0$ and $n_e=1$.  The position of these fixed points is shown in Figure\ \ref{fig:limitcycle}, as a hollow square for $n_e=0$ and a filled square for $n_e=1$.  Further, the shuttle is driven by both Johnson noise and the stochastic fluctuations in $\langle c^\dagger c \rangle$, so it is possible for an autonomous stochastic resonance to appear. 

\citet{gang1993stochastic} characterise ASR phenomena by the emergence, and subsequent growth, of a peak in the power spectrum of the system dynamics as noise power increases.  Following this approach, we numerically calculate the power spectrum $S_x(\tilde{\omega})=|x(\tilde{\omega})|^2$ of the phase space coordinates, where $x(\tilde{\omega})$ is the Fourier transform of $\langle \hat{\mathrm{x}}(t)\rangle$. 

Figure \ref{fig:powerspectrum} shows $S_x(\tilde{\omega})$ for different values of $\gamma$. At $\gamma=0$, the system has only one fixed point, so it fluctuates within the zero point motion. A large value of $\gamma$ is associate with longer distance between the fixed points and consequently larger oscillation.  The  power spectrum exhibits a clear peak around $\tilde \omega=2\omega$, which grows with increasing strength of the Johnson noise, $\gamma$, and is consistent with the characterisation of ASR phenomena.  We also calculate the power spectrum of velocity (Appendix E) and find it has $\pi/2$  phase lag relative to the position, indicating a cyclic orbit in phase space.

%  oscillator has stochastic resonance with frequency  $\omega_{SR}\approx2.0\omega$ which is close to the frequency of the half-harmonic potential. The magnitude of $S(\tilde{\omega}_{SR})$  increases by increasing $\gamma$. 
%  We also calculate the power spectrum of velocity (Appendix E) and find it has $\pi/2$ relative phase to position.
%
% e find support for the existence of autonomous stochastic resonance in the shuttle.

Furthermore, in Appendix E we show phase space histograms for different trajectories over 400 cycles. The `crater-like' depression in the centre of many phase space histograms is another evidence for existence of stochastic resonance.

%In this case the work (area swept in phase space) would increase quasi-monotonically and might also get a "ring-like" structure in phase space.  [Power spectrum of trajectories, has a peak? If so plot in the paper, if not...]

%This would manifest as tantheta= p/x like Rappel increasing in time controlled by the Johnson noise strength.

%Furthermore  we look at single trajectories for different value of $\gamma$ and $\kappa$. Figure \ref{fig:limitcycle} shows these single trajectories for 500 cycles. When $\gamma=0$ the trajectory damps to the steady statewhile for $\gamma>0$, even after 500 cycles the trajectory stays outside of the steady state. The graph (?) shows the last 50 cycles of same trajectories and confirms our observation. We conclude from this observation that single trajectories have a stochastic limit-cycle.

%=======================================================================================================
%                                                                                           Work and Power
%========================================================================================================
\section{Work and Power}
In classical thermodynamics, the energy output of a heat engine is split into a useful part (work) and an entropic part (heat), which depend on the process (or path). In quantum machines with high fluctuation in internal energy, such a classification of work and heat is not obvious. A common approach is based on the change of the total energy $dE = \Tr[H d\rho +\rho dH]$, where the first term on the right hand side is heat and the second term is work \cite{Kosloff-open, alicki1979quantum}. This approach is not suitable for an autonomous engine with a time independent Hamiltonian. Also it has been discussed in \cite{Mahler_Work} that this definition of work does not have an operational meaning. 

Here we define work based on its classical counterpart. The classical definition of infinitesimal mechanical work is $d\mathcal{W}=F d\mathrm{x}$, where $F$ is the force on the work sink. Equivalently instentanous power is $\mathcal{P}=F \text{v}$ where v is the velocity. In quantum mechanics, force and velocity can be correlated i.e. $\langle F \text{v} \rangle \ne  \langle F \rangle \langle \text{v} \rangle $. In the following we define work and power in two different approaches: Semiclassical and fully quantum. 

In semiclassical approach, we ignore quantum correlations between force and position (or velocity). The semiclassical work output during the interval $\tau=t_f-t_i$ is
\begin{align}
\label{eq:WorkSC}
\mathcal{W}_{\text{SC}}(\tau)= \int_{t_i}^{t_f} \langle F(t) \rangle d\langle \hat{\mathrm{x}}(t) \rangle,
\end{align}
and the equivalent instantaneous power is
\begin{align}
\label{eq:PowerSC}
\mathcal{P}_\text{SC}(t) =\langle F(t) \rangle \langle \hat{\text{v}}(t)  \rangle,
\end{align}
where $\hat{\text{v}}(t) = \langle \hat{\mathrm{p}}\rangle/M$ is the velocity of the oscillator. 

For the fully quantum approach we include the correlations between $F$ and $\hat{\text{v}}$. Since these operators do not commute we choose the symmetrised operator, $\frac12 ( F \hat{\mathrm{v}} + \hat{\mathrm{v}} F )$. But this choice has a nonzero value for the ground state. To avoid a zero point power, we subtract $\mathcal{P}_0 = \frac12 \langle 0 | F \hat{\mathrm{v}} + \hat{\mathrm{v}} F |0 \rangle$ from symmetric expectation value where $|0\rangle$ is the gground state of the half-harmonic. We then define quantum power as
\begin{align}
\label{eq:PowerQ}
\mathcal{P}_{\text{Q}}(t)=\frac12 \langle F \hat{\mathrm{v}} + \hat{\mathrm{v}} F \rangle - \mathcal{P}_0.
\end{align}

In either approach, we need to calculate the force on the work sink. We first write Heisenberg equations of motion of phase space coordinates
%\footnote{For a full Harmonic potential, equation (\ref{eq:forcedefinition}) can be simplified to $F  = -\frac{\kappa}{2} p$ which agrees with classical physics where this force is the reaction to the viscous force. In the half-harmonic potential, numerical result shows similar relation as long as the oscillator position is far from the origin. Near the origin, the infinite potential cause a larger force.} 
\begin{align}
\label{eq:Heisenberg}
\frac{d \hat{p}}{dt} &= -M \omega^2 \hat{\mathrm{x}} - \kappa \mathcal{D}^\dagger[a] \hat{p} \nonumber \\
\frac{d \hat{\mathrm{x}}}{dt} &= \frac{\hat{\mathrm{p}}}{M} - \kappa \mathcal{D}^\dagger[a] \hat{\mathrm{x}},
\end{align}
where $\mathcal{D}^\dagger[O]A = O^\dagger A O - \frac12 (O^\dagger O A + A O^\dagger O)$ is the adjoint Lindblad super-operator.\footnote{These equations are in rotating wave approximation and the dissipative terms are in both momentum and position. \cite{bowen2015quantum} } We then compute the force from Newton's law as follows:
\begin{align}
M \frac{d^2 \hat{\mathrm{x}}}{dt^2}   &= \frac{d \hat{\mathrm{p}}}{dt} - \kappa M \mathcal{D}^\dagger[a](\frac{d\mathrm{x}}{dt}) \nonumber \\
					&= -M \omega^2 \hat{\mathrm{x}} -2 \kappa \mathcal{D}^\dagger[a]\hat{p}+ \kappa^2 M \mathcal{D}^\dagger[a] \mathcal{D}^\dagger[a]\hat{\mathrm{x}} \nonumber \\
&=F_{\text{osc}} + F_{\text{Cold Bath}}.
\label{eq:Newton}
\end{align}
Considering that the phase coordinates $\hat{\mathrm{p}}/M$ and $\hat{\mathrm{x}}$ are in dimensionless units and in the same order of magnitude (see Figure \ref{fig:limitcycle}), the last term in equation 20 is negligible when $\kappa \ll \omega$  and $\kappa < 1$. The first term is the oscillation force due to the half-harmonic potential, $F_{\text{osc}} = -M\omega^2 \hat{x}$. The second term is the viscous force due to interaction with the cold reservoir. We then write the working force as the negative of the bath force,
\begin{equation}
F_\text{Work}= -F_\text{Cold Bath} = +2 \kappa \mathcal{D}^\dagger[a]\hat{p}.
\label{eq:Force}
\end{equation}

\begin{figure}[t!]
\includegraphics[width=\columnwidth]{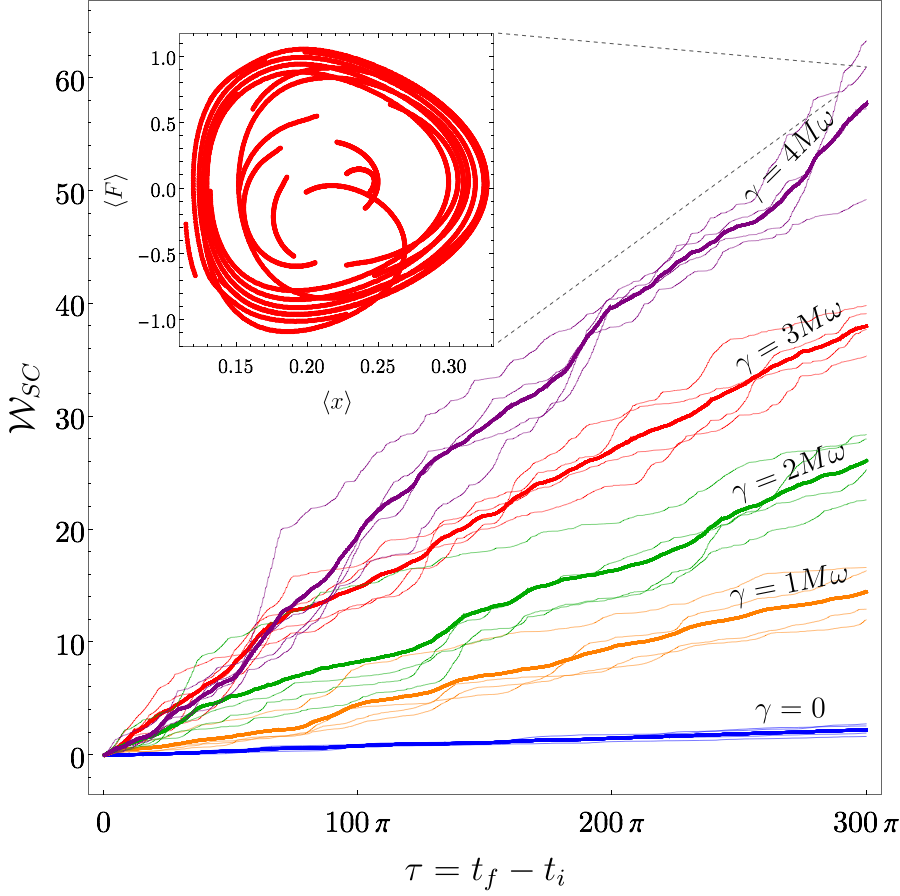}
\caption{Semiclassical work calculated  during the interval $\tau$ for different values of $\gamma$. The interaction rate to the cold reservoir is chosen as $\kappa=0.05$ and initial time, $t_i$ is chosen well after transient to dynamical stationary state. Thin lines are single trajectories while thick lines are averaged over many trajectories. The inset figure shows plot of $\langle F \rangle $ vs. $\langle \hat{\mathrm{x}} \rangle$ for the last 10 cycles of the trajectory with $\gamma = 4 M\omega$. The semiclassical work for each cycle is the area enclosed in the force-position diagram.}
\label{fig:work}
\end{figure}

\begin{figure}[t]
\centering
\includegraphics[width=\columnwidth]{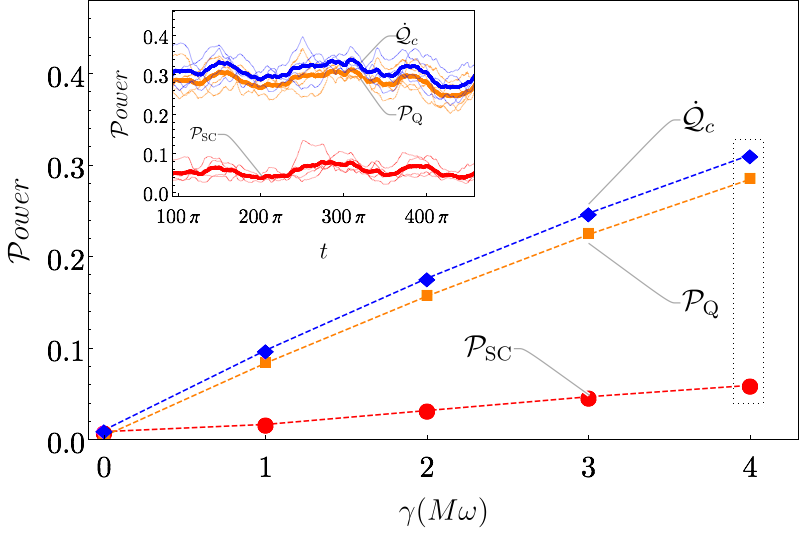}
\caption{Shows time and trajectory averaged of $\mathcal{P}_\text{SC}$ (red circles), $\mathcal{P}_\text{Q}$ (orange squares) and $\dot{\mathcal{Q}}_\text{C}$ (blue diamonds) for different value of $\gamma$ with fixed $\kappa=0.05$. Dashed lines connecting the points to guide the eyes. All the quantities increase by increasing $\gamma$. Quantum power is about 5 times larger than semiclassical which indicates the importance of the force velocity correlation. Also quantum power is close to the heat flux to the cold bath. Inset plot shows $\mathcal{P}_\text{SC}$, $\mathcal{P}_\text{Q}$ and $\dot{\mathcal{Q}}_\text{C}$ for $\gamma=4M\omega$  as functions of time.}
\label{fig:powerplot}
\end{figure}

We use equations (\ref{eq:WorkSC}) and (\ref{eq:Force}) to calculate semiclassical work for different quantum trajectories of the engine. Figure \ref{fig:work} shows $\mathcal{W}_\text{SC}$ for different value of $\gamma$. We have chosen $t_i$ well after all transitions to the dynamical steady state (i.e. system dynamics is independent from the initial state). Thin lines are $\mathcal{W}_\text{SC}$ for single trajectories while thick lines are averaged over trajectories. The inset plot shows the force-position diagram for the last 10 cycles of a single trajectory with $\gamma=4M \omega$. The semiclassical work is the cumulative area enclosed by each cycle of this plot. The work done on the cold reservoir increases linearly (on average) which indicates a constant average power. Red circles in Figure \ref{fig:powerplot} show the averaged semiclassical power. As we expect this power increases with larger $\gamma \propto T_{\text{Hot}}^2$. 

We also calculate power using equation (\ref{eq:PowerQ}) to include quantum correlations. Orange squares in Figure \ref{fig:powerplot} shows $\mathcal{P}_{\text{Q}}$ (averaged over time and over trajectories) for different values of $\gamma$. We see that $\mathcal{P}_{\text{Q}}$ (orange squares) are about 5 times larger than  $\mathcal{P}_{\text{SC}}$ (red circles). The discrepancy between the calculated semiclassical power output, and the fully correlated power output indicates that quantum correlations in the shuttle are significant

Since the cold bosonic reservoir is at zero temperature, we suspect that all the transferred energy into it, is attributable as ``work". To investigate this, we compare $\mathcal{P}_{\text{SC}}$ and $\mathcal{P}_{\text{Q}}$ to $\dot{\mathcal{Q}}_\text{C}$ (blue diamonds in Figure \ref{fig:powerplot}) and we found that quantum power $\mathcal{P}_{\text{Q}} $ is close to $\dot{\mathcal{Q}}_\text{C}$. The small difference between $\mathcal{P}_{\text{Q}}$ and $\dot{\mathcal{Q}}_\text{C}$ is approximately constant. This is shown in the inset plot in Figure \ref{fig:powerplot} which compares time dependent $\mathcal{P}_{\text{SC}}(t)$,$\mathcal{P}_{\text{Q}}(t)$ and $\dot{\mathcal{Q}}_C(t)$ for $\gamma=4M\omega$. This difference is a computational error due to truncation in oscillator energy basis which is more significant in the momentum operator of half-harmonic oscillator that is used to calculate $F_\text{Work}$.
%Although the explanation of the difference between $\mathcal{P}_{\text{Q}}$ and $\dot{\mathcal{Q}}_\text{C}$ is unknown, we hypothesize that it is due to the controller. Checking this hypothesis is not practical using simulated trajectories since the stochastic energetic cost of the controller due to the jump term in equation 15 ($dE_J$) is highly fluctuating and the standard deviation of this noise term is larger than $\dot{\mathcal{Q}}_\text{C} - \mathcal{P}_\text{Q}$.  

The power is affected by the controller implicitly. It can be understood by considering that $\mathcal{P}_\text{Q} \simeq \dot{\mathcal{Q}}_\text{C}$ and $\dot{\mathcal{Q}}_\text{C}$ is proportional to phonon number. Since phonon number is affected by the controller, $\mathcal{P}_\text{Q}$ is affected too.

Our definitions of power were based on the damping of a classical oscillator and we adopted a zero temperature bosonic reservoir as both the cold bath and the work sink. Discriminating heat from work may be done by coupling the engine to a model of a quantum battery that stores the work output as internal energy.  This will be the subject of future work. %\textcolor{red}{We relax this idealization in an upcoming paper, by connecting the engine to a quantum ratchet battery which acts as work sink. Within this operational approach approach we are able to calculate work by the amount of the energy stored in the battery.}

To conclude, we have developed an autonomous heat engine based on the oscillation of a single electron shuttle. In this system the mechanical degree of freedom is the engine and the electric degree of freedom is the controller. We showed that the controller has an energetic cost to the engine power. We also defined both semiclassical and quantum power, and showed that quantum correlation between force and velocity is significant.
%=======================================================================================================
%                                                                                           Acknowledgment
%========================================================================================================
\section*{Acknowledgment}

We would like to thank R. Kosloff, A. Nazir, J. Combes  and C. Muller, K. Modi, F. Pollock and G. J. Milburn for useful discussions. This work was supported by the Australian Research Council Centre of Excellence for Engineered Quantum Systems (Grant No. CE 110001013).

\bibliography{Autonomous}

%=======================================================================================================
%                                                                                           Appendix
%========================================================================================================
\onecolumngrid

\appendix
\newpage
\newpage
\section{Johnson Noise and Master Equation}
Consider an oscillator carries $n_e$ electron and oscillates in an arbitrary potential between two grounded leads with no electric bias. We set one of the leads to be at finite temperature while keeping the other at zero temperature. At finite temperature, the electric potential of the hot lead fluctuates, an effect known as Johnson noise. Consequently an stochastic electric field occurs between the leads. We assume that this stochastic field, which we denote by $\xi(\tau)$, is a white noise and has following average properties
\begin{align}
\langle \xi(\tau) \rangle_{\mathrm{noise}} &= 0 \\
\langle \xi(\tau) \xi(\tau') \rangle_{\mathrm{noise}} &= E^2_{rms} \delta(\tau-\tau')
\end{align}
where $E_{rms}$ is the root mean square of the electric field noise. We assume that the distance between the leads, $d$, is small enough that $E_{rms}$ is constant between the leads and therefore it is proportional to root mean square of Johnson noise as follows
\begin{align}
E^2_{rms}  = \frac{V^2_{rms}}{d} = \frac{4 k_B T R}{d},
\end{align}
where $k_B$ is Boltzmann's constant, $T$ is the temperature of the hot lead, $R$ is the resistance that connects the hot lead to the ground.

The Hamiltonian of the charged oscillator interacting with Johnson noise is
\begin{equation}
\hat{H}_I = H_0 + n_e e \xi(\tau) \hat{\mathrm{x}},
\label{eq:noiseHamiltonian}
\end{equation}
where $H_0$ is the oscillator Hamiltonian.
This system has two different time scales. One is the time scale of Johnson noise which we denote by $\tau$ and the other is the time scale of the oscillator dynamics due to $H_0$ and we label that by $t$. We consider that time scale of noise is much shorter than dynamics, i.e. $d\tau \ll dt$. The master equation of the oscillator after making Born and Markov approximations is
\begin{align}
\dot{\rho}_I =  - (n_e e)^2 \int_0^\infty d\tau' \xi(\tau) \xi(\tau') \left[\hat{\mathrm{x}}_I (\tau), [\hat{\mathrm{x}}_I(\tau'),\rho_I(\tau')] \right]
\end{align}
where subscript $I$ indicates that density matrix $\rho$ is in interaction picture. 

Since we are interested in the dynamics of $H_0$, we average over the noise time scale,
\begin{align}
\dot{\rho}_I & =  - \Big\langle e^2 \int_0^\infty d\tau' \xi(\tau) \xi(\tau') \left[ n_e \hat{\mathrm{x}}_I(\tau) , [ n_e \hat{\mathrm{x}}_I(\tau'),\rho_I(\tau')] \right] \Big \rangle_{\tau} \nonumber\\
		  & =  -  e^2 \int_0^\infty d\tau' \langle \xi(\tau) \xi(\tau') \rangle_{\tau} \left[ n_e \hat{\mathrm{x}}_I(\tau) , [ n_e \hat{\mathrm{x}}_I(\tau'),\rho_I(\tau')] \right] \nonumber \\
		  & =  -  \gamma \int_0^\infty d\tau' \delta(\tau-\tau') \left[ n_e \hat{\mathrm{x}}_I(\tau) , [ n_e \hat{\mathrm{x}}_I(\tau'),\rho_I(\tau')] \right] \nonumber \\
		  & = -  \frac{1}{2} \gamma  \left[  n_e \hat{\mathrm{x}}_I(\tau) , [ n_e \hat{\mathrm{x}}_I(\tau),\rho_I(\tau)] \right]
\end{align}
where $\gamma = e^2 E^2_{rms}$. After expanding the commutator we find
\begin{equation}
\left[ n_e \hat{\mathrm{x}}_I , [ n_e\hat{\mathrm{x}}_I,\rho_I] \right] = -2\mathcal{L}[n_e \hat{\mathrm{x}}_I]\rho_I
\end{equation}
where $\mathcal{L}[O]\rho = O\rho O^\dagger -\frac{1}{2} [O^\dagger O \rho + \rho O^\dagger O]$ is the Lindblad superoperator. Using this result, the master equation in the Schrodinger picture is
\begin{equation}
\dot{\rho} = -i[ H_0 , \rho] + \gamma \mathcal{L}[n_e \hat{\mathrm{x}}]\rho.
\label{eq:masternoise}
\end{equation}

This result shows that the effect of Johnson noise is a dissipative term in the master equation, independent of the shape of the potential. If we include tunnelling on and off the oscillator, we need to substitute  $n_e$ with the electron number $c^\dagger c$ operator in the Hamiltonian (\ref{eq:noiseHamiltonian}), where $c$ is fermionic annihilation operator of an electron. Using the similar derivation we find
\begin{equation}
\dot{\rho} = -i[ H_1 , \rho] + \gamma \mathcal{L}[c^\dagger c \hat{\mathrm{x}}]\rho.
\label{eq:johnson}
\end{equation}
where $H_1$ includes all tunnelling dynamics.

\section{Quantum Mechanics of the Half-Harmonic Potential}
Consider a particle with mass $M$ in a half harmonic potential which is defined as
\begin{equation}
V(x) = \left\{ 
        \begin{array}{ll}
            \frac{1}{2} M  \omega^2 \mathrm{x}^2 & \quad \mathrm{x} \geq 0 \\
            \infty & \quad \mathrm{x} < 0.
        \end{array}
    \right.
\end{equation}    
%Note: we are using units where $\hbar=1$ and $m=1$.\\

The solution to the time-independent Schr\"odinger equation in position space is similar to the full-harmonic oscillator. The boundary condition at $x=0$ enforce the wavefunction to be zero for $x\le0$, therefore only the odd solutions are allowed. The wavefunctions and the energy eigenvalues are
\begin{align}
\psi_n(\mathrm{x}) &= \sqrt{2} \left( \frac{ 1}{\pi} \right)^{\frac{1}{4}} \frac{1}{\sqrt{2^n n!}} \mathrm{H}_n \left( \mathrm{x} \right) e^{-\frac{\mathrm{x}^2}{2}}\\
E_n        &= (n + \frac{1}{2}) \omega ~~~~~ \text{for } n = 1, 3, 5, ...  \quad \text{and } \mathrm{x} \geq 0,
\end{align}
where $\mathrm{H}_n(\mathrm{x})$  are Hermite polynomials and position $\mathrm{x}$ is in units of $x_c = \sqrt{\frac{\hbar}{M \omega}}$. An extra factor of $\sqrt2$ in the wavefunction, in comparison to the full-harmonic oscillator, is due to normalising the wavefunction only in $\mathrm{x}>0$. 

%Also we have written the wavefunction in dimensionless units, using $\psi(x) = \sqrt{x_c} \psi(X)$ and $X=x/x_c$ where tilde used for variables with dimension and the characteristic length is defined as
%\begin{equation}
%x_c = \sqrt{\frac{\hbar}{M \omega}}.
%\end{equation}
We define $| m \rangle$ with $m=0, 1, 2, 3, ... $ as excitations of half-harmonic potential and we write the Hamiltonian eigenequation
\begin{equation}
\label{eq:eigen}
\hat{H} |m\rangle = E_m |m\rangle,
\end{equation}
where $E_m$ is the eigenenergy of the eigenstate $|m\rangle$. We write the odd integers of the half-harmonic, $n$, in terms of $m$, $n=2m+1$. Using this notation, we write the wavefunctions and the energy eigenvalues as 
\begin{align}
\psi_m(\mathrm{x}) &=  \left( \frac{ 1}{\pi} \right)^{\frac{1}{4}} \frac{1}{\sqrt{2^{2m} \cdot (2m+1)!}} H_{2m+1} \left( \mathrm{x} \right) e^{-\frac{\mathrm{x}^2}{2}}\\
E_m        &= (m + \frac{3}{4}) (2 \omega) ~~~~~ \text{for } m=0, 1, 2, 3, ...  \quad \text{and } \mathrm{x} \geq 0. \label{eq:energym}
\end{align}
Furthermore, we define the annihilation, $\hat{a}$, and creation, $\hat{a}^\dagger$, operators of the half-harmonic as follows
\begin{align}
\hat{a} &= \sum_m \sqrt{m+1} |m\rangle \langle m+1| \\
\hat{a}^\dagger &= \sum_m \sqrt{m+1} |m+1\rangle \langle m|.
\end{align}
These operators obey the usual bosonic commutation relation $[\hat{a}, \hat{a}^\dagger] = 1$. Using these operators and equations (\ref{eq:eigen}) and (\ref{eq:energym}), we write the Hamiltonian in terms of the creation and annihilation operators,
\begin{equation}
\label{eq:half_harmonic_hamiltonian}
\hat{H} = \left( \hat{a}^\dagger \hat{a} + \frac{3}{4} \right) (2  \omega).
\end{equation}

This Hamiltonian is similar to the Hamiltonian of a full-harmonic oscillator with the frequency $2\omega$ and the ground state energy, $E_0=\frac{3}{2}\omega$. But there are more differences between half-harmonic and full-harmonic. Specifically, the position and the momentum operators in half-Harmonic are not trivial and there is no simple form of these operators in terms of $\hat{a}$ and $\hat{a}^\dagger$. Instead, we find the matrix elements of these operators in energy basis. 
\begin{equation}
\begin{aligned}
		\hat{\mathrm{x}} &= \sum_{m,m'} |m\rangle \langle m | \hat{\mathrm{x}} |m'\rangle \langle m' | \nonumber \\
				    &=  \sum_{m,m'} \langle m | \hat{\mathrm{x}} |m'\rangle |m\rangle  \langle m' |  \nonumber \\
		   		    &=  \sum_{m,m'} \mathrm{x}_{mm'}  |m\rangle  \langle m' | 
\end{aligned}
\qquad \qquad \qquad
\begin{aligned}
		\hat{\text{p}} &= \sum_{m,m'} |m\rangle \langle m | \hat{\text{p}} |m'\rangle \langle m' | \nonumber \\
				   &=  \sum_{m,m'} \langle m | \hat{\text{p}} |m'\rangle |m\rangle  \langle m' |  \nonumber \\
				   &=  \sum_{m,m'} \text{P}_{mm'}  |m\rangle  \langle m' |
\end{aligned}
\end{equation}
where $ X_{mm'} = \langle m | \hat{\mathrm{x}} |m'\rangle $ and $ P_{mm'} = \langle m | \hat{\text{p}} |m'\rangle $ are matrix elements of $\hat{\mathrm{x}}$ and $\hat{\text{p}}$ and we calculate them, using their known wavefunctions as follows
\begin{equation}
\begin{aligned}
		X_{mm'} = \langle m | \hat{\mathrm{x}} |m'\rangle &= \langle m | \int_0^\infty dx |x\rangle \langle x| \hat{\mathrm{x}} |m'\rangle \nonumber \\
											      &= \langle m | \int_0^\infty  dx  |x\rangle x \langle x | m'\rangle \nonumber \\
											      &= \int_0^\infty  dx  \langle m |x\rangle x \langle x | m'\rangle \nonumber \\
											      &=  \int_0^\infty  \psi_{m}^*(x) x \psi_{m'}(x) dx,
\end{aligned}
\qquad \qquad \qquad
\begin{aligned}
		P_{mm'} = \langle m | \hat{\text{p}} |m'\rangle &= \langle m | \int_0^\infty dx |x\rangle \langle x| \hat{\text{p}} |m'\rangle \nonumber \\
											      &= -i  \langle m | \int_0^\infty  dx  |x\rangle \frac{\partial}{\partial x} \langle x | m'\rangle \nonumber \\
											      &= -i  \int_0^\infty  dx  \langle m |x\rangle \frac{\partial}{\partial x} \langle x | m'\rangle \nonumber \\
											      &= -i  \int_0^\infty  \psi_{m}^*(x) \frac{\partial}{\partial x} \psi_{m'}(x) dx.
\end{aligned}
\end{equation}

\begin{figure*}
  \includegraphics[width=0.8\textwidth]{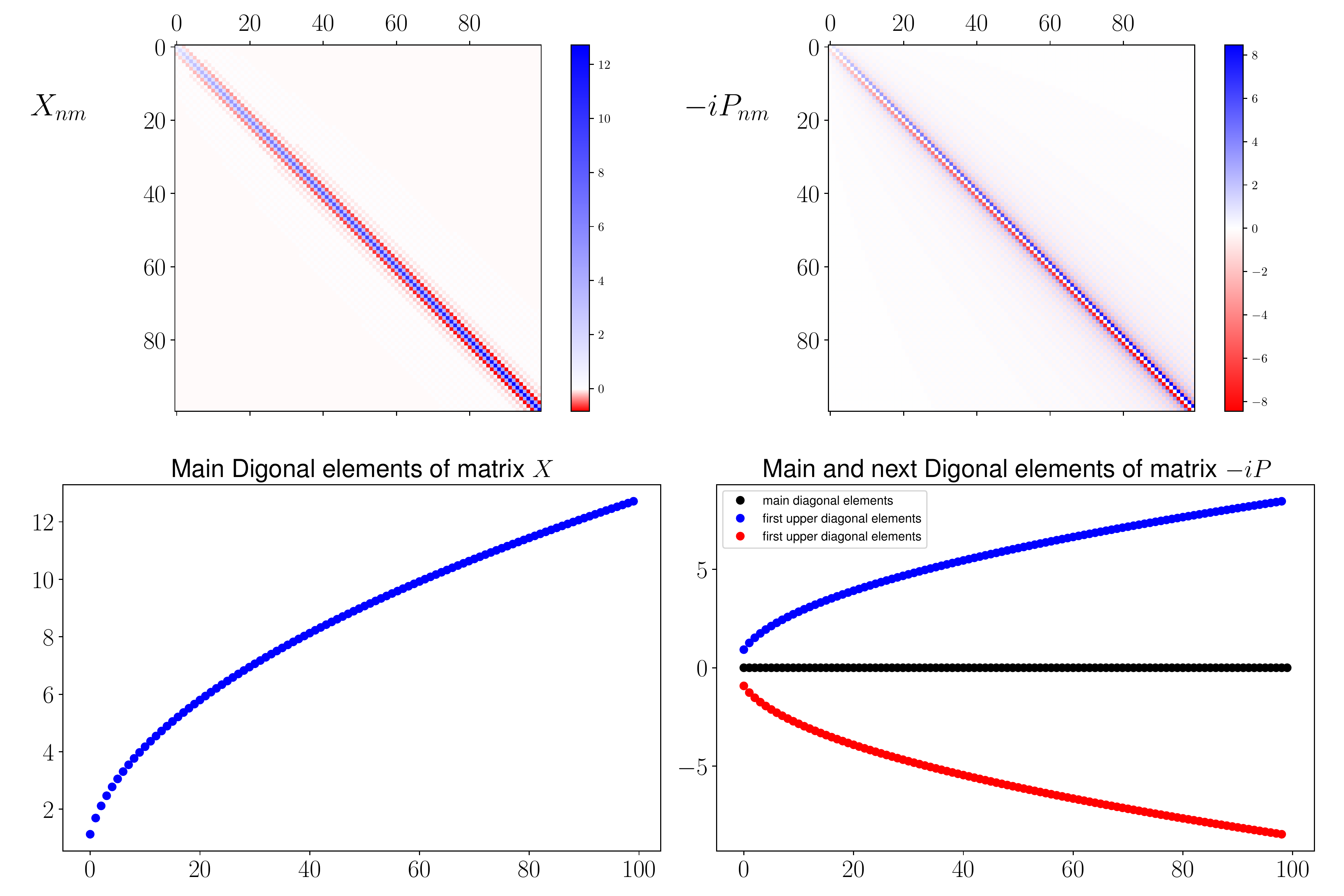}
\caption{shows the matrix elements of the position (a) and momentum (b) operators in the energy eigenbasis. (c) shows the diagonal elements of $X_{nm}$ which are not zero in contrast to the full harmonic potential. (d) shows the main and the first upper and lower diagonal elements of $P_{nm}$}
\label{fig:XandP}
\end{figure*}

Figure \ref{fig:XandP}a and b shows matrix elements of these operators. The diagonal elements of the position operator, Figure \ref{fig:XandP}c, are not zero. This feature is the main reason that half-harmonic potential rectifies Johnson noise. 

These operators obey the canonical commutation relation $[ \hat{\mathrm{x}}, \hat{p} ] = i$, as they must. 

Another interesting property is the commutation relation of the number operator, $a^\dagger a$, with position operator $[ \hat{x}, a^\dagger a] = \frac{i}{2} \hat{p}$ (This commutator in full-harmonic potential is $[\hat{x}, a^\dagger a ] = i\hat{p}$). On the other hand there is no similar commutation relation between $a^\dagger a$ and the momentum operator.

\newpage
\section{Rectifiction of Johnson Noise with the Half-Harmonic Potential}

In this section we show how  a half-harmonic potential rectifies Johnson noise. Consider a charged oscillator with $n_e$ electrons, inside a half-harmonic potential under the effect of Johnson noise and subject to damping with Hamiltonian
\begin{equation}
\label{eq:Hamiltonian_charged}
\begin{split}
\hat{H} &= 2 \hbar \omega \hat{a}^\dagger \hat{a} - e n_e \xi(t) \hat{\mathrm{x}} \\
            &+ \sum_p \hbar \omega_p \hat{d}_p^\dagger \hat{d}_p + \sum_p g(\hat{a}^\dagger \hat{d}_p +\hat{a} \hat{d}_p^\dagger).
\end{split}
\end{equation}
The first term is the Hamiltonian of the half-harmonic oscillator, the second term is due to Johnson noise while the second line shows the hamiltonian of the cold bosonic bath which the oscillator is coupled via the last term. Master equation of the oscillator in Born-Markov approximation is
\begin{equation}
\label{eq:APPmaster_charged}
\dot{\rho} = -i  (2\omega) \left[\hat{a}^\dagger \hat{a}, \rho \right] + \gamma n_e^2 \mathcal{L}[\hat{\mathrm{x}}]\rho + \kappa \mathcal{L}[a]\rho
\end{equation}
where we choose the bosonic bath to be at zero temperature and we used dimensionless units. We write equations of motion for the expectation value of phonon number $N = \langle \hat{a}^\dagger \hat{a} \rangle$
\begin{align}
\label{eq:phonon_motion}
\frac{dN}{dt} = \Tr[ \hat{a}^\dagger \hat{a} \dot{\rho}] =  \frac{n_e^2 \gamma}{4}  - \kappa N
\end{align}
and find the steady state of the phonon number $N_{ss}$ by equating the equation (\ref{eq:phonon_motion}) to zero
\begin{align}
\label{eq:phonon_steadystate}
N_{ss} = \frac{n_e^2 \gamma}{4\kappa}
\end{align}

\begin{figure}[H]
\centering
  \includegraphics[width=0.7\columnwidth]{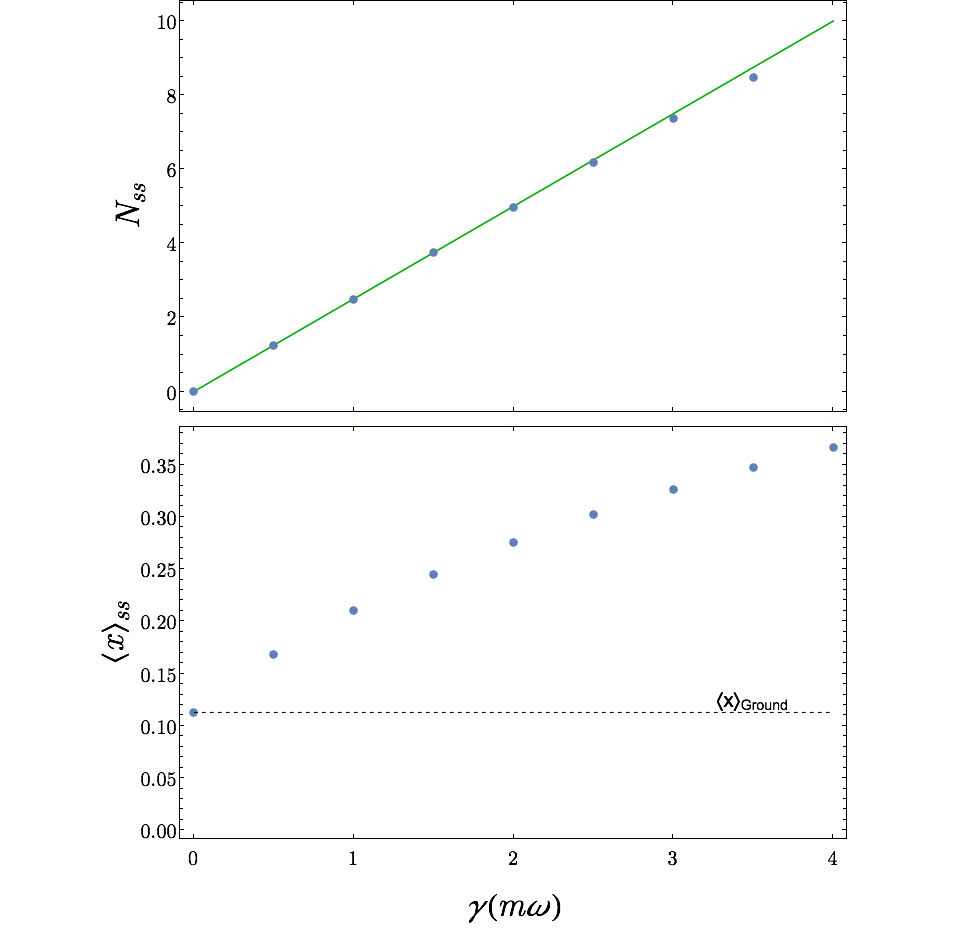} 
    \caption{a) Numerical (blue circles) and analytical (green line) of expectation value of phonon numbers at steady state vs $\gamma$.  b) Numerical calculation of expectation value of position at steady state for differnet $\gamma$. }
\label{fig:SteadyState}
\end{figure}

Equation (\ref{eq:phonon_steadystate}) shows that the average number of phonons at steady state is greater when $\gamma$ (Johnson noise) is larger and $\kappa$ (damping rate) is smaller. This shows that Johnson noise act as an energy source (hot bath) and bosonic reservoir acts as an energy sink (cold bath). Also the average number of phonons is zero when $n_e = 0$. This means that an uncharged shuttle looses all mechanical energy to the cold bath. Figure \ref{fig:SteadyState}a shows $N_{ss}$ as a function of $\gamma$ for a fixed $\kappa$. Numerical calculations of steady states for phonon numbers (blue circles) matches with the analytical result of equation (\ref{eq:phonon_steadystate}) (green line).

To show the rectification of the half-harmonic potential we need to show that Johnson noise affects the steady state of the position. In a full-harmonic potential, the steady state of the position is zero even after being heated by Johnson noise. Due to the complication in the position operator of the half-harmonic potential, writing an equation of motion for the position is impossible. Alternatively, we derive the steady state of the master equation numerically and then calculate the expectation value of the position operator. Figure \ref{fig:SteadyState}b shows the expectation value of position at steady state for different value $\gamma$. At $\gamma=0$, the steady state of position is as it is for the ground state. Larger $\gamma$ causes an increase in $\langle \mathrm{x} \rangle_{ss}$. This result shows that Johnson noise applies, on average, a net force on the charged shuttle and we conclude that the half-harmonic potential rectifies the Johnson noise. 

In addition we show that the steady states are stable points (i.e. attractors or sinks). To see this, we solve the master equation (\ref{eq:master_charged}) with an initial state near to the steady states. Specifically, we chose the steady state and displaced it by a small amount. We found that with this choice of initial state the oscillator damps to the related steady state. Figure \ref{fig:SteadyState} shows unconditional solutions to the master equation (\ref{eq:master_charged}) with different initial conditions (blue), each near to a steady state (dashed black line). All the solutions damp to their related steady states. 

\begin{figure}[h!]
  \includegraphics[width=0.6\columnwidth]{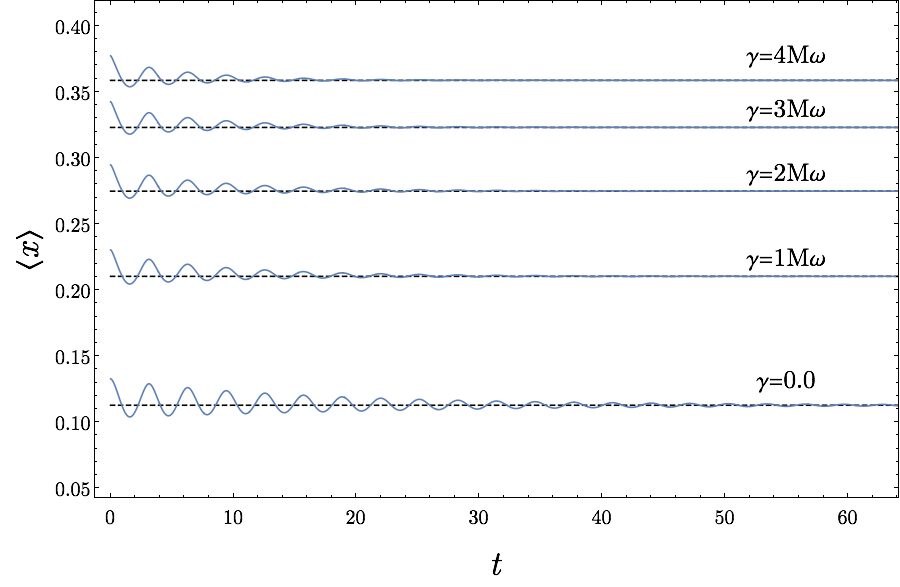} 
    \caption{Unconditional solutions to the master equation near steady state for different values of $\gamma$. Each solution eventually damps to the steady state.}
\label{fig:NearSS}
\end{figure}

\newpage
\section{Derivation of the Master Equation}
We showed the effect of Johnson noise after average over the noise and found that adding Johnson noise to the Hamiltonian will add the term due to Johnson noise (the last term of equation (\ref{eq:johnson}) ) to the master equation. In this section we consider an electron shuttle without Johnson noise and find its master equation in the Born-Markov approximation and at the end add the Johnson noise term to the master equation.
 
The Hamiltonian of the shuttle, the leads and the bosonic reservoir (for damping) is
\begin{align}
H_1 &= H_0 + H_I \\
H_0 &= \hbar\omega \hat{a}^\dagger \hat{a} + \hbar\omega_I\hat{c}^\dagger\hat{c} + \sum_{\ell,k}\omega_k \hat{b}_{\ell,k}^\dagger\hat{b}_{\ell,k} + \sum_p \hbar \omega_p \hat{d}_p^\dagger \hat{d}_p\\
H_I  &= \sum_{\ell,k} \tau_{\ell,k} F_\ell(\hat{\mathrm{x}}) \hat{b}_{\ell,k} \hat{c}^\dagger + \sum_p g_p \hat{a}^\dagger \hat{d}_p + \mathrm{h.c.}
\end{align}

We first move to the interaction frame
\begin{align}
H_I(t) &= e^{iH_0t/\hbar} H_I e^{-iH_0t/\hbar} \\
          &= \sum_{\ell,k}\tau_{\ell k} \left(e^{i\omega \hat{a}^\dagger \hat{a}t} F_\ell(\hat{\mathrm{x}}) e^{-i\omega \hat{a}^\dagger \hat{a}t}\right) 
                                             \left( e^{i \omega_I \hat{c}^\dagger \hat{c} t}\hat{c}^\dagger e^{-i \omega_I \hat{c}^\dagger \hat{c} t} \right)
                                             \left(e^{i \sum_{\ell,k}\omega_k\hat{b}_{\ell k}^\dagger\hat{b}_{\ell k}t}\hat{b}_{\ell k} e^{- i \sum_{\ell,k}\omega_k\hat{b}_{\ell k}^\dagger\hat{b}_{\ell k}t}\right) \nonumber\\
          &+ \sum_p \left( e^{i \omega \hat{a}^\dagger \hat{a} t} \hat{a}^\dagger e^{-i\omega \hat{a}^\dagger \hat{a} t} \right) \left( g_p e^{i\sum_p \omega_p \hat{d}_p^\dagger \hat{d}_p t} \hat{d}_p e^{-i\sum_p \omega_p \hat{d}_p^\dagger \hat{d}_p t}\right) +\mathrm{h.c.}  \\
          &=  \sum_{\ell,k}\tau_{\ell k}\tilde{F}_\ell(\hat{\mathrm{x}}) ~\hat{c}^\dagger e^{i\omega_I t} ~ \hat{b}_{\ell k} e^{-i\omega_kt} + \hat{a}^\dagger e^{i\omega t} \sum_p g_p \hat{d}_p e^{-i\omega_p t}+ \mathrm{h.c.}
\end{align}
where the time dependence of $H_I(t)$ shows we are in the interaction frame and $\tilde{F}_\ell(\hat{\mathrm{x}}) = e^{i\omega \hat{a}^\dagger \hat{a}t} F_\ell(\hat{\mathrm{x}}) e^{-i\omega \hat{a}^\dagger \hat{a}t}$. Then we decompose the system and bath operator by defining
\begin{subequations}
\begin{align}
C_{1\ell}(t) &\equiv \tilde{F}_\ell(\hat{\mathrm{x}}) \hat{c}^\dagger e^{i\omega_I t} & 
C_{2\ell}(t) &\equiv \tilde{F}_\ell(\hat{\mathrm{x}}) \hat{c} e^{-i\omega_I t} \\
B_{1\ell}(t) &\equiv  \sum_k \tau_{\ell k} \hat{b}_{\ell k} e^{-i\omega_kt} &
B_{2\ell}(t) &\equiv  \sum_k \tau_{\ell k} \hat{b}_{\ell k}^\dagger e^{i\omega_kt}
\end{align}
\end{subequations}

\begin{subequations}
\begin{align}
A_1(t) &\equiv \hat{a}^\dagger e^{i\omega t} & 
A_2(t) &\equiv \hat{a} e^{-i\omega t}\\
D_1(t) &\equiv \sum_p g_p \hat{d}_p e^{-i\omega_p t} &
D_2(t) &\equiv \sum_p g_p \hat{d}^\dagger_p e^{i\omega_p t}
\end{align}
\end{subequations}
and we write the interaction Hamiltonian in terms of these operators
\begin{equation}
H_I(t) = \sum_{i,\ell} C_{i\ell}(t) \otimes B_{i\ell}(t) + A_i(t) \otimes D_i(t).
\end{equation} 

Now the master equation becomes
\begin{align}
\dot{\rho}(t) = &-\sum_{\ell=s,d} \sum_{i,j} \int_0^\infty d\tau \left(\left[ C_{i\ell}(t), C_{j\ell}(t-\tau) \rho(t) \right] \mathcal{B}_{ij}^\ell(\tau) + \left[ \rho(t) C_{j\ell}(t-\tau),C_{i\ell}(t)\right] \mathcal{B}_{ji}^\ell(-\tau) \right) \nonumber \\
                      &-\sum_{i,j} \int_0^\infty d\tau \left(\left[ A_i(t), A_j(t-\tau) \rho(t) \right] \mathcal{D}_{ij}(\tau) + \left[ \rho(t) A_j(t-\tau),A_i(t)\right] \mathcal{D}_{ji}(-\tau) \right)
\end{align}
where $\mathcal{B}_{ij}^\ell(t-t')=\langle B_{i\ell}(t) B_{j\ell}(t') \rangle_E$ and $\mathcal{D}_{ij}(t-t') = \langle D_{i}(t) D_{j}(t') \rangle_E$ are baths correlations and to calculate them we use the following averages,

\begin{subequations}
\begin{align}
\langle b_{\ell k} b_{lk'} \rangle_E                             &= \Tr[b_{\ell k} b_{lk'} \rho_E]                            = 0                                              &  \langle d_p d_{p'} \rangle_E                             &= \Tr[d_p d_{p'} \rho_E]                               = 0\\
\langle b_{\ell k}^\dagger b_{lk'}^\dagger \rangle_E &=\Tr[b_{\ell k}^\dagger b_{lk'}^\dagger \rho_E] = 0                                               &  \langle d_p^\dagger d_{p'}^\dagger \rangle_E &=\Tr[d_p^\dagger d_{p'}^\dagger \rho_E]    = 0   \\
\langle b_{\ell k}^\dagger b_{lk'} \rangle_E               &=\Tr[b_{\ell k}^\dagger b_{lk'} \rho_E]               = \delta_{kk'} f_\ell(\omega_k)        &  \langle d_p^\dagger d_{p'} \rangle_E               &=\Tr[d_p^\dagger d_{p'} \rho_E]                  = \delta_{pp'} n(\omega_p)\\
\langle b_{\ell k} b_{lk'}^\dagger \rangle_E               &= \Tr[b_{\ell k} b_{lk'}^\dagger \rho_E]              = \delta_{kk'} (1-f_\ell(\omega_k) )  &  \langle d_p d_{p'}^\dagger \rangle_E               &= \Tr[d_p d_{p'}^\dagger \rho_E]                = \delta_{pp'} (1+n(\omega_p) )
\end{align}
\end{subequations}
where $f_{\ell}(\omega_I) = 1/(e^{(\omega_I - \mu)/T_{\ell}}+1)$ is Fermi distribution of lead $\ell$ and $n(\omega_p) = 1/(e^{\omega_p/T} - 1)$ is the Bose-Einestien distribution for the cold reservoir. Using these we find

\begin{subequations}
\begin{align}
\mathcal{B}_{11}^\ell(t-t') = \langle B_{1\ell}(t) B_{1\ell}(t') \rangle_E  &= \langle B_{1\ell}(t-t') B_{1\ell}(0) \rangle_E = 0\\
\mathcal{B}_{22}^\ell(t-t') = \langle B_{2\ell}(t) B_{2\ell}(t') \rangle_E  &= \langle B_{2\ell}(t-t') B_{2\ell}(0) \rangle_E = 0 \\
\mathcal{B}_{12}^\ell(t-t') = \langle B_{1\ell}(t) B_{2\ell}(t') \rangle_E  &=\langle B_{1\ell}(t-t') B_{2\ell}(0) \rangle_E \nonumber\\
                                                          &=  \sum_{kk'} \tau_{\ell k} \tau_{lk'} \langle \hat{b}_{\ell k}  \hat{b}_{lk'}^\dagger \rangle_E e^{-i\omega_k(t-t')} \nonumber \\
                                                          &=  \sum_{k} \tau_{\ell k}^2  (1-f_\ell(\omega_k) ) e^{-i\omega_k(t-t')}  \nonumber \\
                                                          &=   \int_0^\infty d\epsilon J_\ell(\epsilon) (1-f_\ell(\epsilon)) e^{-i\epsilon (t-t')} \\
\mathcal{B}_{21}^\ell(t-t') = \langle B_{2l}(t) B_{1\ell}(t') \rangle_E  &=\langle B_{2\ell}(t-t') B_{1\ell}(0) \rangle_E \nonumber\\
                                                          &= \sum_{kk'} \tau_{\ell k} \tau_{lk'} \langle \hat{b}_{\ell k}^\dagger  \hat{b}_{lk'} \rangle_E e^{i\omega_k(t-t')} \nonumber \\
                                                          &=  \sum_{k} \tau_{\ell k}^2  f_\ell(\omega_k) e^{i\omega_k(t-t')}  \nonumber \\
                                                          &=   \int_0^\infty d\epsilon J_\ell(\epsilon) f_\ell(\epsilon) e^{i\epsilon (t-t')}
\end{align}
\end{subequations}
where we have defined the spectral density $J_\ell(\epsilon) \equiv \sum_k |\tau_{\ell k}|^2 \delta(\epsilon - \omega_k)$. Similarly we find the bosonic bath correlation
\begin{subequations}
\begin{align}
\mathcal{D}_{11}(t-t') = \langle D_{1}(t) D_{1}(t') \rangle_E  &= \langle D_{1}(t-t') D_{1} \rangle_E = 0\\
\mathcal{D}_{22}(t-t') = \langle D_{2}(t) D_{2}(t') \rangle_E  &= \langle D_{2}(t-t') D_{2} \rangle_E = 0 \\
\mathcal{D}_{12}(t-t') = \langle D_{1}(t) D_{2}(t') \rangle_E  &=\langle D_{1}(t-t') D_{2} \rangle_E \nonumber\\
                                                          &=  \sum_{pp'} g_p g_{p'} \langle \hat{d}_{d}  \hat{d}_{p'}^\dagger \rangle_E e^{-i\omega_{p}(t-t')} \nonumber \\
                                                          &=  \sum_{p} g_p^2  (1+n(\omega_p) ) e^{-i\omega_{p}(t-t')} \nonumber \\
                                                          &= \int_0^\infty d\epsilon J_b(\epsilon) (1+n(\epsilon)) e^{-i\epsilon (t-t')}\\
\mathcal{D}_{21}(t-t') = \langle D_{2}(t) D_{1}(t') \rangle_E  &=\langle D_{2}(t-t') D_{1} \rangle_E \nonumber\\
                                                          &=  \sum_{pp'} g_p g_{p'} \langle \hat{b}_{p}^\dagger  \hat{b}_{p'} \rangle_E e^{i\omega_{p}(t-t')} \nonumber \\
                                                          &=  \sum_{p} g_p^2  n(\omega_p) e^{i\omega_{p}(t-t')}  \nonumber \\
                                                          &= \int_0^\infty d\omega J_b(\epsilon) n(\epsilon) e^{i\epsilon (t-t')}
\end{align}
\end{subequations}
where we defined spectral density of bosonic bath as $J_b(\epsilon) \equiv \sum_k |g_p|^2 \delta(\epsilon - \omega_p)$. Using these correlation functions, we write the master equation as sum of integrals,
\begin{equation}
\dot{\rho}(t) = - \sum_{\ell=s,d} \sum_{i=1}^4 \mathcal{I}_i^\ell - \sum_{i=1}^4 \mathcal{K}_i
\end{equation}
where $\mathcal{I}_i^\ell$s are integrals for lead $\ell$ as follows 
\begin{align}
\mathcal{I}_1^\ell &= \int_0^\infty d\tau \left[ C_{1\ell}(t), C_{2\ell}(t-\tau) \rho(t) \right] \mathcal{B}_{12}^\ell(\tau) \nonumber \\
                      &=\int_0^\infty d\tau \left[\tilde{F}_\ell\hat{c}^\dagger e^{i\omega_I t} , \tilde{F}_\ell\hat{c} e^{-i\omega_I(t-\tau)} \rho(t)\right]  \int_0^\infty d\epsilon J_\ell(\epsilon) (1-f_\ell(\epsilon)) e^{-i\epsilon \tau} \nonumber \\
                      &=  [\tilde{F}_\ell \hat{c}^\dagger, \tilde{F}_\ell\hat{c} \rho(t)] \int_0^\infty d\tau \int_0^\infty d\epsilon J(\epsilon) (1-f_\ell(\epsilon)) e^{i(\omega_I - \epsilon) \tau}\\
                      &=  [\tilde{F}_\ell \hat{c}^\dagger, \tilde{F}_\ell \hat{c} \rho(t)] \frac{\Gamma_\ell(\omega_I)}{2}(1-f_\ell(\omega_I))
\end{align}

\begin{align}
\mathcal{I}_2^\ell &= \int_0^\infty d\tau \left[\rho(t) C_{2\ell}(t-\tau), C_{1\ell}(t) \right] \mathcal{B}_{21}^\ell(-\tau) \nonumber \\
                      &=\int_0^\infty d\tau \left[\rho(t) \tilde{F}_\ell \hat{c} e^{-i\omega_I(t-\tau)} , \tilde{F}_\ell \hat{c}^\dagger e^{i\omega_I t}\right] \int_0^\infty d\epsilon J_\ell(\epsilon) f_\ell(\epsilon) e^{-i\epsilon \tau} \nonumber \\
                      &= [\rho(t)\tilde{F}_\ell \hat{c}, \tilde{F}_\ell \hat{c}^\dagger] \int_0^\infty d\tau \int_0^\infty d\epsilon J_\ell(\epsilon) f_\ell(\epsilon) e^{i(\omega_I-\epsilon) \tau}\\
                      &=  [\rho(t)\tilde{F}_\ell \hat{c}, \tilde{F}_\ell \hat{c}^\dagger]\frac{\Gamma_\ell(\omega_I)}{2} f_\ell(\omega_I)
\end{align}

\begin{align}
\mathcal{I}_3^\ell &= \int_0^\infty d\tau \left[ C_{2\ell}(t), C_{1\ell}(t-\tau) \rho(t) \right] \mathcal{B}_{21}^\ell(\tau) \nonumber \\
                      &=\int_0^\infty d\tau \left[\tilde{F}_\ell \hat{c} e^{-i\omega_I t} , \tilde{F}_\ell \hat{c}^\dagger e^{i\omega_I(t-\tau)} \rho(t)\right] \int_0^\infty d\epsilon J_\ell(\epsilon) f_\ell(\epsilon) e^{i\epsilon \tau} \nonumber \\
                      &= [\tilde{F}_\ell \hat{c}, \tilde{F}_\ell \hat{c}^\dagger \rho(t)] \int_0^\infty d\tau \int_0^\infty d\epsilon J_\ell(\epsilon) f_\ell(\epsilon) e^{-i(\omega_I-\epsilon) \tau}\\
                      &= [\tilde{F}_\ell \hat{c}, \tilde{F}_\ell \hat{c}^\dagger \rho(t)] \frac{\Gamma_\ell(\omega_I)}{2} f_\ell(\omega_I)
\end{align}

\begin{align}
\mathcal{I}_4^\ell &= \int_0^\infty d\tau \left[\rho(t) C_{1\ell}(t-\tau), C_{2\ell}(t) \right] \mathcal{B}_{12}^\ell(-\tau) \nonumber \\
                      &=\int_0^\infty d\tau \left[\rho(t) \tilde{F}_\ell \hat{c}^\dagger e^{i\omega_I(t-\tau)}, \tilde{F}_\ell \hat{c} e^{-i\omega_I t}\right] \int_0^\infty d\epsilon J_\ell(\epsilon) (1-f_\ell(\epsilon)) e^{i\epsilon \tau} \nonumber \\
                      &= [\rho(t) \tilde{F}_\ell \hat{c}^\dagger, \tilde{F}_\ell \hat{c}] \int_0^\infty d\tau \int_0^\infty d\epsilon J_\ell(\epsilon) (1-f_\ell(\epsilon)) e^{-i(\omega_I-\epsilon) \tau}\\
                      &= [\rho(t) \tilde{F}_\ell \hat{c}^\dagger, \tilde{F}_\ell \hat{c}] \frac{\Gamma_\ell(\omega_I)}{2} (1-f_\ell(\omega_I))
\end{align}
and $\mathcal{K}_i$ are integrals for bosonic bath,
\begin{align}
\mathcal{K}_1 &= \int_0^\infty d\tau \left[ A_1(t), A_2(t-\tau) \rho(t) \right] \mathcal{D}_{12}(\tau) \nonumber \\
                      &=\int_0^\infty d\tau \left[\hat{a}^\dagger e^{i\omega t} , \hat{a} e^{-i\omega(t-\tau)} \rho(t)\right] \int_0^\infty d\epsilon J_b(\epsilon) (1+n(\epsilon)) e^{-i\epsilon \tau} \nonumber \\
                      &= [\hat{a}^\dagger, \hat{a} \rho(t)] \int_0^\infty d\tau \int_0^\infty d\epsilon J_b(\epsilon) (1+n(\epsilon)) e^{i(\omega-\epsilon) \tau}\\
                      &= [\hat{a}^\dagger, \hat{a} \rho(t)] \frac{\kappa(\omega)}{2} (1+n(\omega))
\end{align}

\begin{align}
\mathcal{K}_2 &= \int_0^\infty d\tau \left[\rho(t) A_2(t-\tau), A_1(t) \right] \mathcal{D}_{21}(-\tau) \nonumber \\
                      &=\int_0^\infty d\tau \left[\rho(t) \hat{a} e^{-i\omega(t-\tau)} , \hat{a}^\dagger e^{i\omega t}\right] \int_0^\infty d\epsilon J_b(\epsilon) n(\epsilon) e^{-i\epsilon \tau} \nonumber \\
                      &= [\rho(t)\hat{a}, \hat{a}^\dagger] \int_0^\infty d\tau \int_0^\infty d\epsilon J_b(\epsilon) n(\epsilon) e^{i(\omega-\epsilon) \tau}\\
                      &= [\rho(t)\hat{a}, \hat{a}^\dagger] \frac{\kappa(\omega)}{2} n(\omega)
\end{align}

\begin{align}
\mathcal{K}_3 &= \int_0^\infty d\tau \left[ A_2(t), A_1(t-\tau) \rho(t) \right] \mathcal{D}_{21}(\tau) \nonumber \\
                      &=\int_0^\infty d\tau \left[\hat{a} e^{-i\omega t} , \hat{a}^\dagger e^{i\omega(t-\tau)} \rho(t)\right] \int_0^\infty d\epsilon J_b(\epsilon) n(\epsilon) e^{i\epsilon \tau} \nonumber \\
                      &= [\hat{a}, \hat{a}^\dagger \rho(t)] \int_0^\infty d\tau \int_0^\infty d\epsilon J_b(\epsilon) n(\epsilon) e^{-i(\omega-\epsilon) \tau}\\
                      &= [\hat{a}, \hat{a}^\dagger \rho(t)] \frac{\kappa(\omega)}{2} n(\omega)
\end{align}

\begin{align}
\mathcal{K}_4 &= \int_0^\infty d\tau \left[\rho(t) A_1(t-\tau), A_2(t) \right] \mathcal{D}_{12}(-\tau) \nonumber \\
                      &=\int_0^\infty d\tau \left[\rho(t) \hat{a}^\dagger e^{i\omega(t-\tau)}, \hat{a} e^{-i\omega t}\right] \int_0^\infty d\epsilon J_b(\epsilon) (1+n(\epsilon)) e^{i\epsilon \tau} \nonumber \\
                      &= [\rho(t) \hat{a}^\dagger, \hat{a}] \int_0^\infty d\tau \int_0^\infty d\epsilon J_b(\epsilon) (1+n(\epsilon)) e^{-i(\omega-\epsilon) \tau}\\
                      &= [\rho(t) \hat{a}^\dagger, \hat{a}] \frac{\kappa(\omega)}{2} (1+n(\omega))
\end{align}
We used the following relation to calculate the time integrals
\begin{equation}
\int_0^\infty dt e^{\pm i \epsilon t} = \pi \delta(\epsilon) \pm i \frac{P}{\epsilon} ,
\end{equation}
where $P$ denotes the Cauchy principal value. We only use the real part of the integral since the imaginary part can be absorbed into $H_0$. Also we defind the following zero temperature rates

\begin{align}
\Gamma_\ell(\epsilon) &= 2 \pi J_\ell(\epsilon) \\
\kappa(\epsilon) &= 2 \pi J_b(\epsilon).
\end{align} 

Now we write the master equation by adding the calculated integrals
\begin{align}
\dot{\rho}(t)  &= - \sum_{\ell=s,d}  &&\left( \frac{\Gamma_\ell}{2} (1-f_\ell(\omega_I))  [\tilde{F}_\ell \hat{c}^\dagger, \tilde{F}_\ell \hat{c} \rho(t)] +\frac{\Gamma_\ell}{2} f_\ell(\omega_I) [\rho(t) \tilde{F}_\ell \hat{c}, \tilde{F}_\ell \hat{c}^\dagger] \right. \nonumber \\
                    & &+& \left.  \frac{\Gamma_\ell}{2} f_\ell(\omega_I)  [\tilde{F}_\ell \hat{c}, \tilde{F}_\ell \hat{c}^\dagger \rho(t)] +  \frac{\Gamma_\ell}{2} (1-f_\ell(\omega_I)) [\rho(t) \tilde{F}_\ell \hat{c}^\dagger, \tilde{F}_\ell \hat{c}] \right) \nonumber \\
                    &  &-& \left( \frac{\kappa}{2} (1+n(\omega)) [\hat{a}^\dagger, \hat{a} \rho(t)] + \frac{\kappa}{2} n(\omega) [\rho(t)\hat{a}, \hat{a}^\dagger] \right. \nonumber \\
                    & &+& \left. \frac{\kappa}{2} n(\omega) [\hat{a}, \hat{a}^\dagger \rho(t)] + \frac{\kappa}{2} (1+n(\omega))  [\rho(t) \hat{a}^\dagger, \hat{a}]  \right) \nonumber \\
                    &=- \sum_{\ell=s,d}  &&\left( \frac{\Gamma_\ell}{2} (1-f_\ell(\omega_I)) \left( [\tilde{F}_\ell \hat{c}^\dagger, \tilde{F}_\ell \hat{c} \rho(t)] +  [\rho(t) \tilde{F}_\ell \hat{c}^\dagger, \tilde{F}_\ell \hat{c}]  \right) + \frac{\Gamma_\ell}{2} f_\ell(\omega_I) \left( [\rho(t)\tilde{F}_\ell \hat{c}, \tilde{F}_\ell \hat{c}^\dagger] +  [\tilde{F}_\ell \hat{c}, \tilde{F}_\ell \hat{c}^\dagger \rho(t)] \right) \right) \nonumber \\
                    & &-&\left( \frac{\kappa}{2} (1+n(\omega)) \left( [\hat{a}^\dagger, \hat{a} \rho(t)] +  [\rho(t) \hat{a}^\dagger, \hat{a}] \right) + \frac{\kappa}{2} n(\omega) \left([\rho(t)\hat{a}, \hat{a}^\dagger] +  [\hat{a}, \hat{a}^\dagger \rho(t)]  \right) \right) \nonumber\\
                    &= \sum_{\ell=s,d}  &&\left( \Gamma_\ell (1-f_\ell(\omega_I))\mathcal{L}[\tilde{F}_\ell \hat{c}]\rho + \Gamma_\ell f_\ell(\omega_I) \mathcal{L}[\tilde{F}_\ell \hat{c}^\dagger]\rho \right) + \left( \kappa (1+n(\omega)) \mathcal{L}[\hat{a}]\rho + \kappa n(\omega)\mathcal{L}[\hat{a}^\dagger] \rho \right).
\end{align}

And the master equation in its final form is

\begin{align}
\dot{\rho}(t)  &= \Gamma_s (1-f_s(\omega_I))\mathcal{L}[\tilde{F}_s \hat{c}]\rho + \Gamma_s f_s(\omega_I) \mathcal{L}[\tilde{F}_s \hat{c}^\dagger]\rho \nonumber \\
                    &+ \Gamma_d (1-f_d(\omega_I))\mathcal{L}[\tilde{F}_d \hat{c}]\rho + \Gamma_\ell f_d(\omega_I) \mathcal{L}[\tilde{F}_d \hat{c}^\dagger]\rho \nonumber \\
                    &+ \kappa (1+n(\omega)) \mathcal{L}[\hat{a}]\rho + \kappa n(\omega)\mathcal{L}[\hat{a}^\dagger] \rho
\end{align}

\newpage
\section{Evidence for Stochastic Resonance}
In this section we show the evidence for existence of Stochastic Resonance.

The following table of histograms shows the momentum distribution of solutions to the conditional master equation. The value of $\gamma$ is increasing from top to bottom and each histogram at each row is a different trajectory for the same parameters. Gaussian fit shown in blue curve and the residual is shown below each histogram.

Gaussian fit matches pretty well and show the oscillator is mostly in thermal equilibrium. In the middle of histograms, some disagreements with the gaussian fit are seen. These are due to non equilibrium state of the oscillator after electron jumps. 
\begin{figure}[h]
\centering
\includegraphics[width=\columnwidth,height=0.8\textheight]{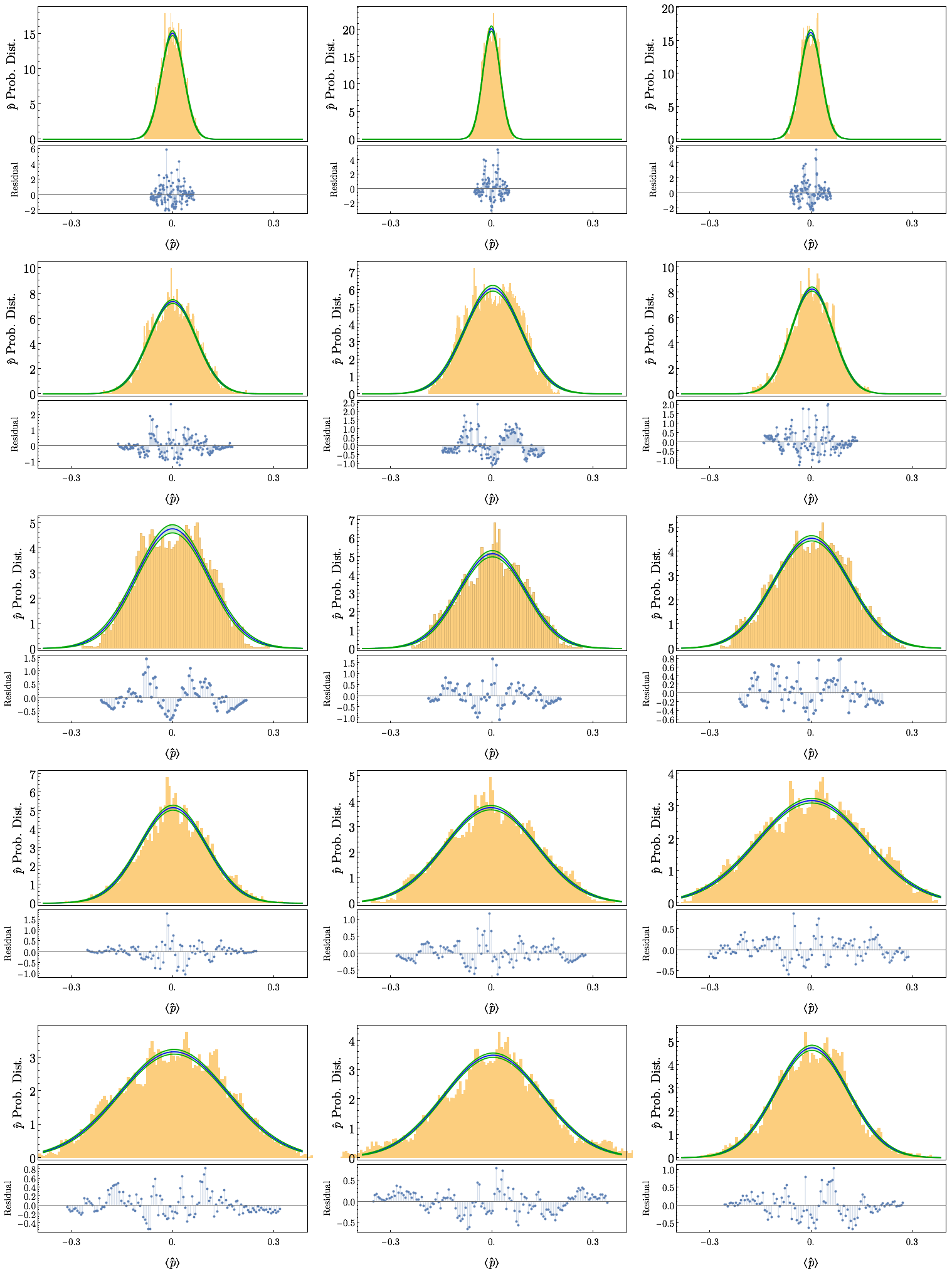}
\label{fig:histograms}
\end{figure}

~

~

The following 3D histograms show distribution of the phase space coordinates over 400 cycles. Similar to the previous table of histograms, the value of $\gamma$ increases from top to bottom and each histogram at each row represents a different trajectory for the same parameters. The `crater-like' depression in the centre of these histograms suggests the existence of a stochastic resonance.

\begin{figure}[h]
\centering
\includegraphics[width=\columnwidth,height=0.85\textheight]{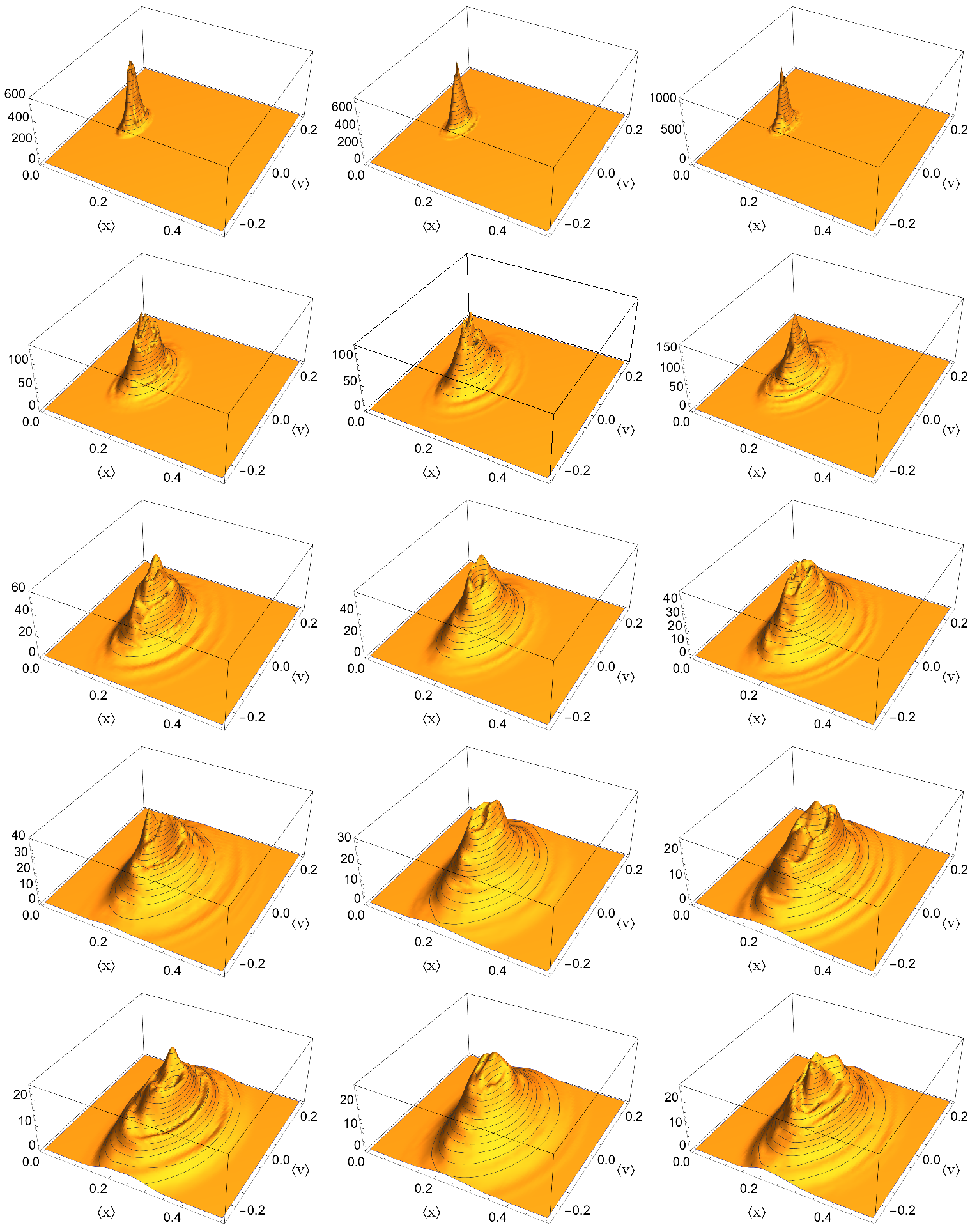}
\label{fig:phasehistplots}
\end{figure}

Following figures show the power spectrum of the position ($S_x(\tilde{\omega})$) and the velocity ($S_v(\tilde{\omega})$). The peak at $\tilde \omega \approx 2.0 \omega$ shows stochastic resonance frequency.
\begin{figure}[h]
\centering
\includegraphics[width=0.9\columnwidth]{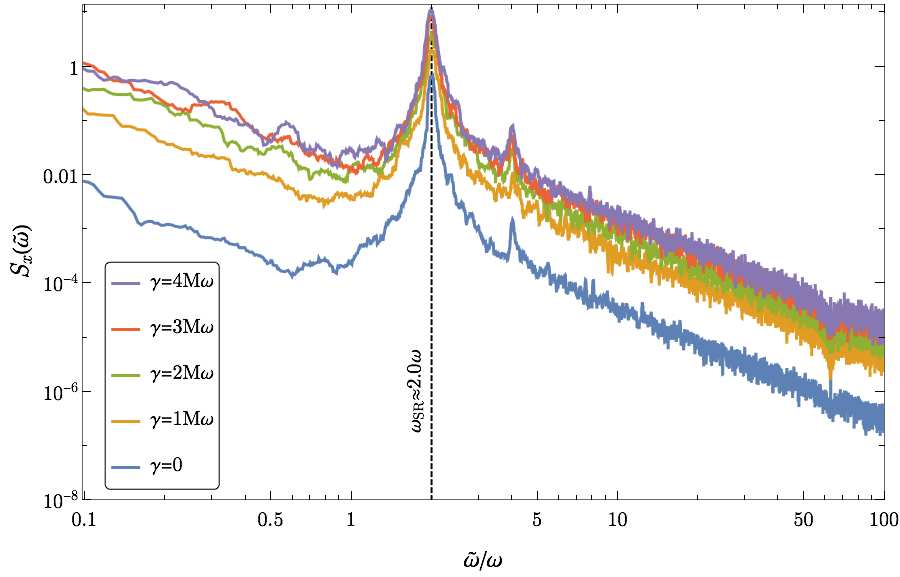}
\includegraphics[width=0.9\columnwidth]{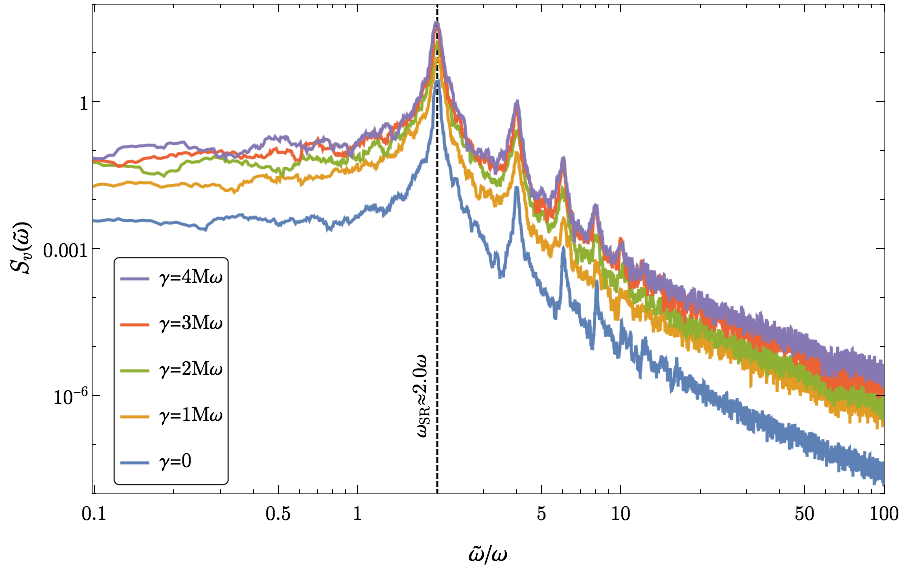}
\label{fig:SLogPlot}
\end{figure}

\end{document}